\documentclass[aps,pra,showpacs,twocolumn,superscriptaddress]{revtex4-1}
\usepackage{amssymb,latexsym,mathrsfs}
\usepackage{amsfonts}
\usepackage{graphicx}
\usepackage{lmodern}
\usepackage{fix-cm}
\usepackage{xfrac}
\usepackage{color}    
\usepackage{textcomp}
\usepackage{amsmath}
\usepackage{subfigure}
\usepackage{float}
\usepackage{datetime}
\usepackage{cancel}
\usepackage{hyperref}
\hypersetup{
    linkcolor=red,
    filecolor=magenta,      
    urlcolor=cyan,
}
\usepackage{tabularx,ragged2e,booktabs}

\newcommand{\bearr}{\begin{array}}
\newcommand{\enarr}{\end{array}}

\newcommand{\bea}{\begin{eqnarray}}
\newcommand{\eea}{\end{eqnarray}}
\def \be{\begin{eqnarray}}
\def \ee{\end{eqnarray}}

\begin{document}

\title{Multilayered density profile for noninteracting fermions in a rotating two-dimensional~trap}

\author{Manas Kulkarni}
\affiliation{International Centre for Theoretical Sciences, Tata Institute of Fundamental Research, Bengaluru -- 560089, India}

\author{Satya N. Majumdar}
\affiliation{LPTMS, CNRS, Univ.  Paris-Sud,  Universit\'e Paris-Saclay,  91405 Orsay,  France}

\author{Gr\'egory Schehr}
\affiliation{LPTMS, CNRS, Univ.  Paris-Sud,  Universit\'e Paris-Saclay,  91405 Orsay,  France}
\affiliation{Sorbonne Universit\'e, Laboratoire de Physique Th\'eorique et Hautes Energies, CNRS UMR 7589, 4 Place Jussieu, 75252 Paris Cedex 05, France}

\date{\today}
\begin{abstract}
We compute exactly the average spatial density for $N$ spinless noninteracting fermions in a $2d$ harmonic 
trap rotating with a constant frequency $\Omega$ in the presence of an additional repulsive central potential $\gamma/r^2$. We find that, 
in the large $N$ limit, the bulk density has a rich and nontrivial profile -- with a hole at the center of the trap and
surrounded by a multi-layered ``wedding cake'' structure. The number of layers depends on $N$ and on the two parameters 
$\Omega$ and $\gamma$ leading to a rich phase diagram. Zooming in on the edge of the $k^{\rm th}$ layer, we find
that the edge density profile exhibits $k$ kinks located at the zeroes of the $k^{\rm th}$ Hermite polynomial. 
Interestingly, in the large $k$ limit, we show that the edge density profile approaches a limiting form, which resembles 
the shape of a propagating front, found in the unitary evolution of certain quantum spin chains. We also study how a newly formed droplet 
grows in size on top of the last layer as one changes the parameters. 

\end{abstract}

\setcounter{page}{1}
\maketitle
\section{Introduction}
Noninteracting spinless fermions in a confining trap is a subject of much current theoretical and
experimental interest \cite{bloch2008many,nascimbene2010equation,cheuk2015quantum,haller2015single,parsons2015site,mukherjee2017homogeneous,hueck2018two}. On one hand, this system is realisable in cold atom experiments, and several techniques such as 
absorption imaging \cite{inguscio2008ultra, giorgini2008theory, joseph2011observation} for collective density measurements and quantum gas microscopes \cite{qm1,qm2,qm3} for direct in situ imaging of the individual fermions with remarkably high resolutions are available. On the 
other hand, it is simple enough to be analytically tractable and yet exhibits rich and nontrivial spatial fluctuations, 
even at zero temperature, due to the Pauli exclusion principle \cite{vicari2012entanglement,eisler2013universality,marino2014phase,dean2015finite,dean2015universal,dean2016noninteracting,marino2016number}. While the bulk density is usually well described by the 
local density approximation (LDA) \cite{butts1997trapped,inguscio2008ultra}, this approximation breaks down near the edges of the Fermi gas, induced by the trap.
A number of recent studies have pointed out that LDA is not sufficient to capture the density fluctuations and correlations near the edges \cite{kohn1998edge,eisler2013universality,dean2015finite,dean2015universal,dean2016noninteracting}.
For certain one-dimensional trapping potentials, such as the harmonic trap, an exact mapping was found between the positions
of the fermions in the ground state and the eigenvalues of a suitable random matrix ensemble \cite{eisler2013universality,marino2014phase} -- for a recent review see \cite{dean2019noninteracting}. Using results from
the random matrix theory (RMT), the density correlations near the edges were computed exactly and their universal
properties (with respect to the shape of the trapping potential) were elucidated \cite{eisler2013universality,dean2015finite,dean2015universal,calabrese2015random,dean2016noninteracting,lacroix2017statistics,dean2018wigner,le2018multicritical,cunden2019free}. The connection to RMT does not
hold generically in higher dimensions. However, using the determinantal properties of the noninteracting fermions, 
the edge properties in higher dimensions could still be computed analytically \cite{dean2015universal,dean2016noninteracting}. 
   
A particularly interesting situation corresponds to fermions in a rotating trap in two-dimensions, which has been
studied recently both experimentally~\cite{ho2000rapidly,aftalion2005vortex,schweikhard2004rapidly} and theoretically~\cite{fetter2009rotating, cooper2008rapidly,lacroix2019rotating}. In this system, the single particle Hamiltonian, in the rotating frame, 
is given by \cite{landau1980statistical,leggett2006quantum}
\begin{equation}
\label{ham_intro}
\hat H = \frac{{p}^2}{2m} + V(r) - \Omega{L}_z
\end{equation}
where $V(r)$ is a confining central potential,  
${L}_z= x p_y- y p_x = -i  (x \partial_y - y \partial_x)$
is the $z$-component of the angular momentum and $\Omega$ is
the rotation frequency. For the harmonic trap $V(r) = (1/2) m \omega^2 r^2$, 
an important parameter is the ratio $\nu = \Omega/\omega$, which must 
satisfy $0<\nu<1$ to keep the fermions confined. The limit $\nu \to 0$ corresponds
to fermions in a non-rotating harmonic trap while in the opposite limit $\nu \to 1$,
this problem can be mapped to the celebrated Landau problem of noninteracting
fermions in a plane in the presence of a perpendicular magnetic field \cite{cooper2012quantum}. Interestingly, in this
$\nu \to 1$ limit, the positions of $N$ fermions in the ground state map onto the eigenvalues
of the classical complex Ginibre ensemble of RMT \cite{lacroix2019rotating}, where one considers a random 
$N \times N$ matrix with independent complex Gaussian entries \cite{forrester2010log}. In this mapping, one assumes
that the $N$ fermions are confined in the lowest Landau level, which can be realized by setting
$1 - 2/N < \nu < 1$. With this assumption, the bulk density for large $N$ is rather simple: it is just uniform 
over the disk of radius $\sqrt{N}$ centred at the origin~\cite{lacroix2019rotating}. 


\begin{figure}[t]
\textbf{a)} \includegraphics[width=.85\linewidth]{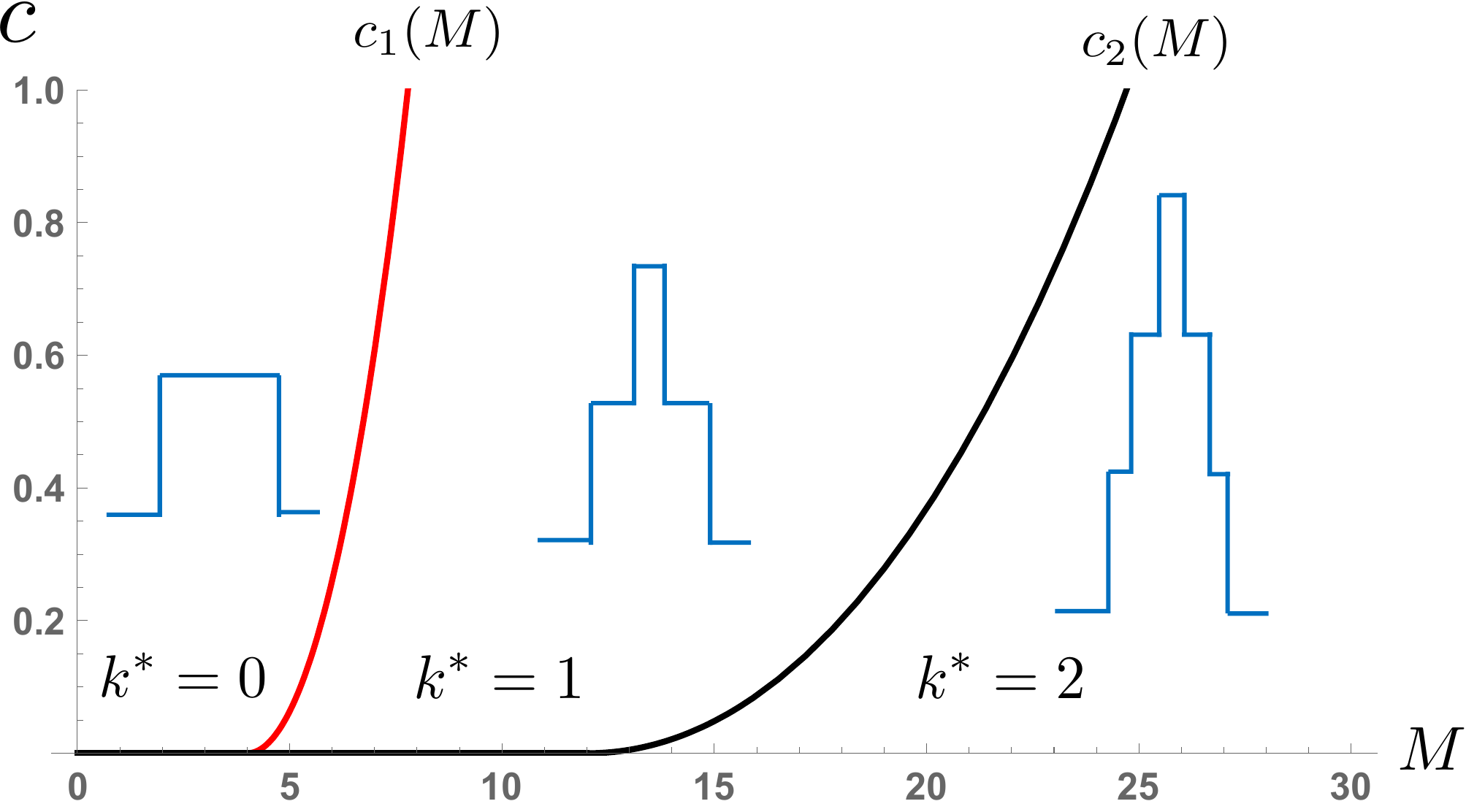}\hfill
\\ \textbf{b)} \includegraphics[width=.85\linewidth]{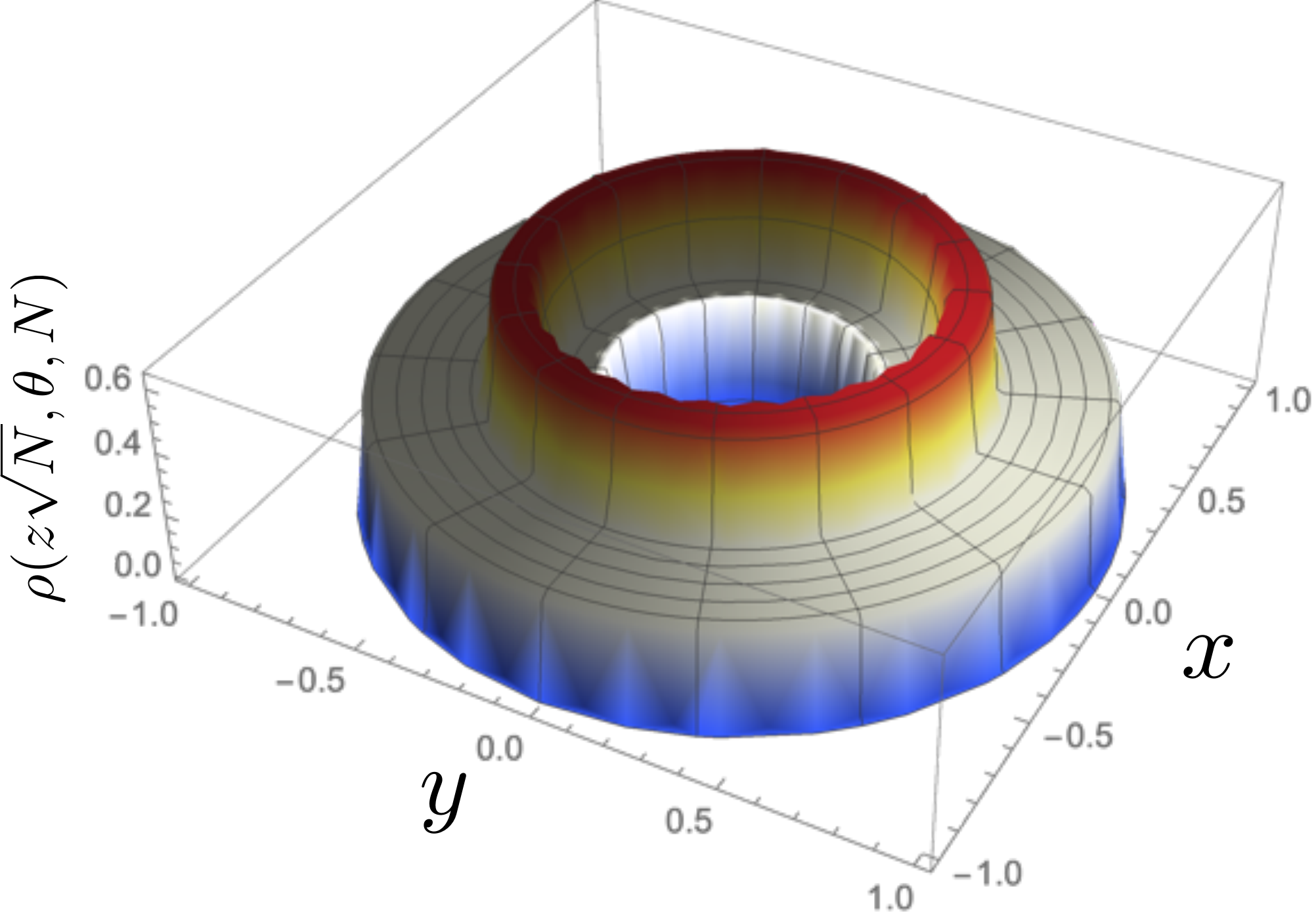}
\hfill\\ \textbf{c)}
\includegraphics[width=.85\linewidth]{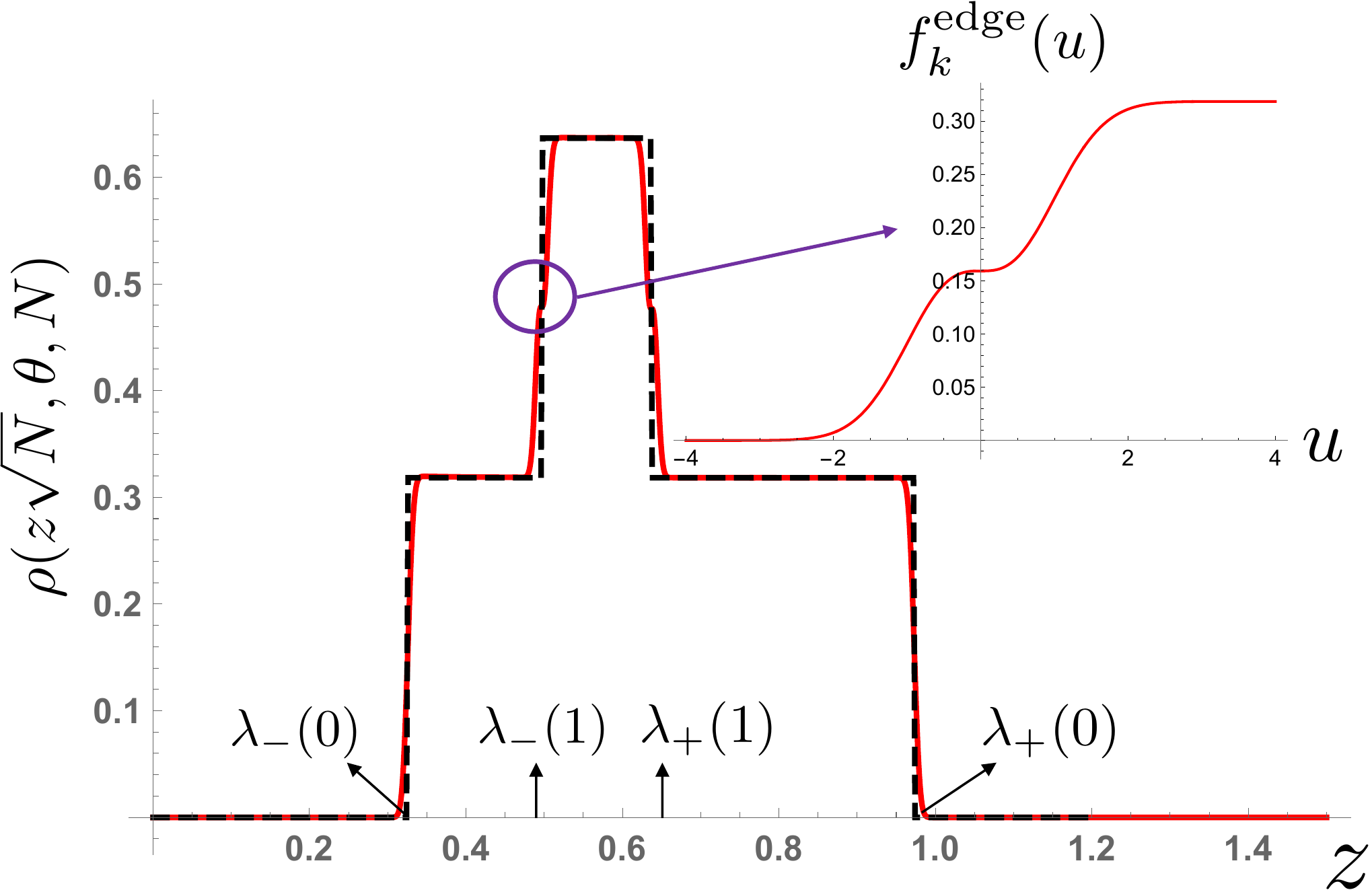}
\caption{{\bf (a)}: Phase diagram in the $(M,c)$ plane. It is divided
into regions labeled by $k^* =0, 1, 2...$ denoting the number of bands
($n$) that are below the Fermi level. The lines $c_{n}(M)$ separates
the regions between $k^*=n-1$ and $k^*=n$. In each of the regions, a
typical (representative) density profile is shown (blue). We see that
every new band creates a new layer in the density. {\bf (b)}: A 3D
representation of the exact density in (\ref{gen_density}). A hole
around the origin is surrounded by a multi-layered ``wedding cake''
structure. {\bf (c)}: Plot showing the comparison between the exact
density in \eqref{gen_density} (red solid) and the large $N$
asymptotic bulk density \eqref{rho_bulk} (black dashed) for $c=1$,
$M=10$ and $N=8000$ (this corresponds to $k^*=1$ in the phase
diagram). We zoomed in on the left edge of the $k=1$ layer and the
inset shows the scaling function $f^{\rm edge}_1(u)$ in
(\ref{rho_edge}) plotted vs $u$.}\label{combinedFig}
\end{figure}

This uniform bulk density emerges because, in the ground state, the fermions are all in the lowest Landau
level. A natural question then is: how the density may change if the many-body ground state also contains single-particle 
states belonging to higher Landau levels? Indeed, this is a generic situation as one increases $N$, for fixed $\nu$. 
In addition, since the potential $V(r)$ is radially symmetric, it is convenient to solve the corresponding Schr\"odinger 
equation in polar coordinates, which will automatically generate an effective repulsive interaction $\sim 1/r^2$ in the radial
direction. Hence it is natural to consider a more generic potential from the start 
\begin{eqnarray} \label{def_V}
V(r) = \frac{1}{2} m\omega^2 r^2 + \frac{\gamma}{2r^2} \;, \;\; \gamma \geq 0 \;.
\end{eqnarray}
We thus have two parameters $0<\nu<1$ and $\gamma \geq 0$. In this 
paper we investigate the density profile
in the ground state, for large $N$, as a function of $\nu$ and $\gamma$ and find an extremely rich phase
diagram in the $(\nu, \gamma)$ plane.

Let us first summarise our main results. We find that in the large $N$ limit the appropriate
rescaled parameters are 
\begin{eqnarray} \label{def_cM}
c = \frac{\gamma}{N} \; \;\; {\rm and} \;\;\; M = (1-\nu^2)\,N \;,
\end{eqnarray} 
which are both kept of order $O(1)$ as $N \to \infty$. We will show later that this scaling is
necessary to keep the average density of fermions of order $O(1)$ as $N \to \infty$. The phase diagram in 
the $(M,c)$ plane is depicted in Fig. \ref{combinedFig} a). There are series of critical lines $c_1(M), c_2(M), \cdots$
that separate the regions labelled by $k^*$ where $k^*+1$ is the number of Landau levels included in the ground state. 
As one crosses these critical lines, the density profile undergoes abrupt changes, as shown in Fig. \ref{combinedFig} a).
For a given $k^*$ the bulk density vanishes for $r<  \sqrt{l_-(0)}$ thus creating a hole around the origin [see Fig. \ref{combinedFig} b)].   
Outside the hole, the density is nonzero over an annulus $\sqrt{l_-(0)}< r < \sqrt{l_+(0)}$. On top of this annulus, there is a 
``wedding cake'' structure [see Fig. \ref{combinedFig} b)] with $k^*$ layers with progressively smaller supports but with equal heights $1/\pi$. For example the $k$-th layer has support on $\sqrt{l_-(k)}< r < \sqrt{l_+(k)}$ (see Fig.~\ref{combinedFig}). As shown later, $l_{\pm}(k) = O(N)$. We also investigated the change in the density profile as one crosses the critical lines in the phase diagram and found an interesting ``travelling front structure'' in the density. 
Furthermore, if we zoom in on the left boundary of the $k$-th layer (and symmetrically on the right boundary), i.e., close to $\sqrt{l_{-}(k)}$ (and symmetrically at $\sqrt{l_+(k)}$) we find a nontrivial edge-profile of the density (\ref{rho_edge}) with $k$ kinks whose locations coincide with the zeros of the $k$-th Hermite polynomial $H_k(-u) = 0$ with $u$ denoting the scaled distance from $\sqrt{l_-(k)}$ (see inset of Fig. \ref{combinedFig} (c)). Finally, in the limit where $k \gg 1$, the edge profile approaches a nontrivial limiting form, which we compute exactly. Interestingly, the same limiting form has appeared in completely different problems, such as in a propagating one-dimensional fermionic front separating a high and low density phases and  
evolving unitarily in time~\cite{antal1999transport,antal2008logarithmic,eisler2013full,hunyadi2004dynamic,mukherjee2018quantum}.

\section{Model and Properties}

\begin{figure}[t]
\centering
    \includegraphics[width=.95\linewidth]{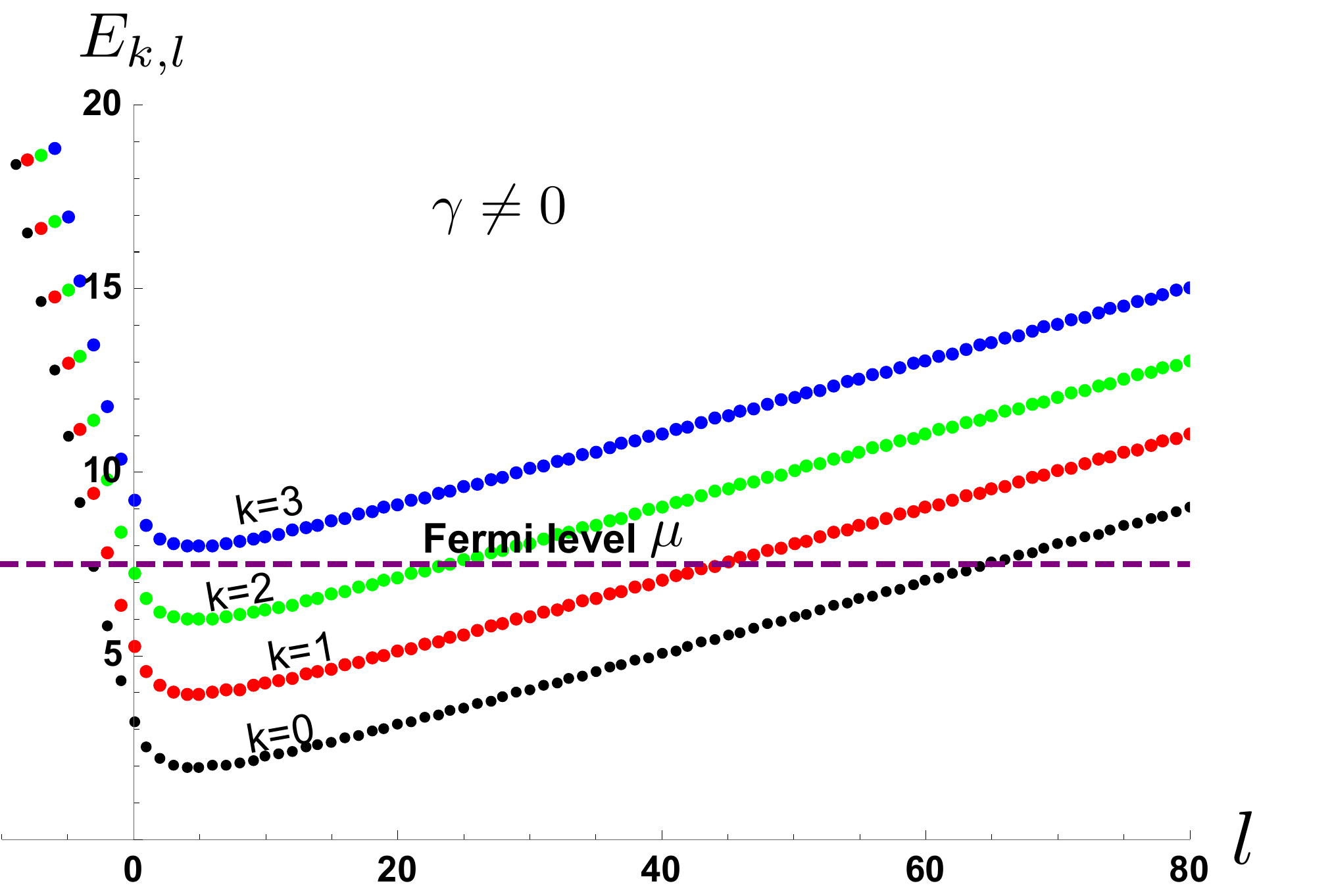}
    \caption{Energy levels $E_{k,l}$ in \eqref{eq:meigenvaluesg} vs $l$ for $k=0, 1, 2, 3$, for $\gamma = 5$ and $\nu=0.9$. The purple (dashed) horizontal line marks the Fermi level, $\mu = 7.5$. Only the states with energy below $\mu$ can contribute to the ground state.}\label{fig:mdisp}
\end{figure}

We start with the single particle Hamiltonian in~(\ref{ham_intro}) with $V(r)$ in Eq.~(\ref{def_V}). The model turns out to be integrable in the sense that the
Schr\"odinger equation $\hat H \psi_{k,l}(r,\theta) = E_{k,l} \psi_{k,l}(r,\theta)$ is exactly solvable in the polar coordinates (see Appendix \ref{sec:mp} for details). For convenience, we set $m=\hbar = 1$. We get  
\begin{equation}
\psi_{k,l}(r,\theta) = a_{k,l} L_k^{\lambda}(r^2) r^\lambda e^{-r^2/2} e^{i l \theta} \;, \; {\rm with} \;\; \lambda = \sqrt{\gamma + l^2} \;,
\label{eq:meigf}
\end{equation}
where $L_{k}^{\lambda}(x)$ are the generalised Laguerre polynomials and the normalisation gives 
$a_{k,l}^2 = \frac{\Gamma(k+1)}{\pi \Gamma(k+1 + \lambda)}$. The associated eigenvalues, in units of $\omega$, are given by~(see Appendix \ref{sec:elev} for details)
\begin{equation}
E_{k,l}= 2k+1+\sqrt{\gamma+l^2}- \nu l \;.
\label{eq:meigenvaluesg}
\end{equation}
The single particle states are labelled by a pair of integers $(k,l)$
with $k=0,1,2...$ and $l=0, \pm 1, \pm 2, ...$. The energy levels (\ref{eq:meigenvaluesg}) are shown in Fig.~\ref{fig:mdisp}. Different values of $k$ correspond to different bands or Landau levels. 

We now consider $N$ spinless noninteracting fermions in their ground state. The many-body ground-state is thus given by a Slater determinant 
constructed from $N$ single particle eigenfunctions associated to the lowest $N$ eigenvalues. For a given $N$, the eigenfunctions participating in the 
Slater determinant may belong to multiple bands of the spectrum in Fig. \ref{fig:mdisp} with $k^*$ denoting the label of the highest band which is at least partially filled. We also denote by $\mu$ the Fermi energy, i.e. the energy of the highest occupied single particle energy level. The Fermi energy $\mu$ can
be tuned by varying $N$. As $\mu$ increases, one sees from Fig. \ref{fig:mdisp}, where $\mu$ is indicated by a horizontal line, that more and more states with energy levels below $\mu$ contribute to the ground state since $k^*$ also increases.

The average number density, normalised to $N$, at a point ${\bf r}=(r,\theta)$ is given by $\rho(r,\theta,N) = \sum_{i=1}^N \langle \delta({\bf r} - {\bf r_i}) \rangle$ where $\langle \cdots \rangle$ denotes the expectation value in the ground state. For noninteracting fermions, it can be computed explicitly in terms of single particle eigenfunctions
\begin{eqnarray} \label{gen_density}
\rho(r, \theta,N) = \sum_{k,l} |\psi_{k,l}(r,\theta)|^2= \sum_{k=0}^{k^*} \rho_k(r,\theta, N) \;,
\end{eqnarray}
where $\rho_k(r, \theta, N)$ denotes the density from the $k^{\text{th}}$ band and is given by
\begin{equation} \label{rho_k}
\rho_k(r,\theta, N) = \frac{\Gamma(k+1)\,e^{-r^2}}{\pi} \sum_{l = l_-(k)}^{l_+(k)} \frac{[L_k^\lambda(r^2)]^2 \; r^{2 \lambda}}{\Gamma(\lambda+k+1)} \;.
\end{equation}
Here, $l_{\mp}(k)$ are the locations where the Fermi level $\mu$ intersects the $k^{\text{th}}$ band, i.e. $E_{k,l_{\pm}(k)} = \mu$. Solving this equation using (\ref{eq:meigenvaluesg}) (shifting energy by 1, effectively absorbing it in $\mu$) gives  
\begin{equation} \label{lpmk_1}
l_{\pm}(k) = \frac{\nu(\mu - 2k) \pm \sqrt{(\mu-2k)^2 - \gamma(1-\nu^2)}}{1-\nu^2} \;.
\end{equation} 
For a given $k$, $E_{k,l}$ (Eq.~\ref{eq:meigenvaluesg}) has a minimum at $l=l^*$ where $l^* = \frac{\nu}{\sqrt{1-\nu^2}} \sqrt{\gamma} $. 
Note that $l^*$ is independent of $k$ and the energy of the $k^{\text{th}}$ band at this minimum is given by $E_{k, l^*} = 2 k + \sqrt{(1-\nu^2)\gamma} $. 
If the Fermi level has to intersect at least one band, we must have $E_{0,l^\star}<\mu$ which implies $\mu > \sqrt{(1-\nu^2) \gamma}$.
For a fixed $\mu$, the number of bands $k^*$ below $\mu$ can be obtained by simultaneously requiring
$E_{k^*,l^\star} = 2k^* + \sqrt{(1-\nu^2)\gamma}  < \mu $ and $ E_{k^*+1,l^\star} = 2k^* + 2 + \sqrt{(1-\nu^2)\gamma}  > \mu $
which yields $k^* = \rm{Int}\left[ {\frac {\mu - \sqrt{(1-\nu^2)\gamma}}{2}} \right] $ where ${\rm Int}(x)$ denotes the integer part of $x$ (see Appendix \ref{sec:elev} for details). The relation between the Fermi energy $\mu$ and $N$ can be obtained by counting the total number of single particle levels with energy below $\mu$. This gives $\sum_{k=0}^{k^*}\left[ l_+(k) - l_-(k) \right] = N$ which fixes $\mu$ in terms of $N$ (see Appendix \ref{sec:elev} for details). For large $N$, it turns out that $\mu \sim O(1)$.  
%
%
So far, the results are exact for arbitrary $N$, $\nu \in (0,1)$ and $\gamma>0$. To make further progress we now work in the large $N$  
limit and re-scale the two parameters $\nu$ and $\gamma$ as in Eq. (\ref{def_cM}). Thus $c$ and $M$ are the new rescaled parameters. 
It turns out that this scaling is necessary to keep $\rho(r,\theta,N) = O(1)$ for large $N$. In terms of $c$ and $M$, we have from Eq.~(\ref{lpmk_1}),
$l_{\pm}(k) = \lambda_{\pm}(k) N $ where $\lambda_{\pm}(k) = \frac{(\mu - 2 k) \pm \sqrt{(\mu-2k)^2 - c M}}{M}$ and $k^* = {\rm Int}\left[ \frac{\mu - \sqrt{cM}}{2}\right]$. 

\section{Critical lines in the $(M,c)$ plane} 

The $(M,c)$ plane is divided into different regions labeled by $k^* = 0, 1, 2, \ldots$ separated by critical lines [see Fig. \ref{combinedFig} a)].   For there to be $k^*$ bands, we require, $2 k^* + \sqrt{cM} < \mu < 2 (k^*+1) + \sqrt{c M}$. Setting the upper bound,  $\mu = 2 (k^*+1) + \sqrt{c M}$, one gets $\frac{4}{M}\sum_{q=1}^{k^*+1} \sqrt{q(q+\sqrt{cM})} = 1 $~(see Appendix \ref{sec:cmp} for details). Solving this equation for $c$ as a function of $M$ gives the critical line $c_{k^*+1}(M)$. Thus, in the $(M,c)$ plane, we get different regions labeled by $k^* = 0, 1, 2 \ldots$. The region between $c_{n}(M)$ and $c_{n+1}(M)$ corresponds to the region with $k^* = n$, i.e. the Fermi level includes $n$ bands below it. For instance, $c_1(M)$ and $c_2(M)$ can be explicitly computed~(see Appendix \ref{sec:cmp} for details) and are plotted in Fig.~\ref{combinedFig} a), e.g., $c_1(M) = (1/M) \left( {M^2}/{16}-1\right)^2 \Theta(M-4) $ where $ \Theta(x) $ is the Heaviside step function. \\



\section{Density in the large-N limit} 
We start by analysing the large $N$ limit of $\rho_k(r,\theta,N)$ in Eq.~(\ref{rho_k}) upon setting $r = z \sqrt{N}$. Since $l_{\pm}(k) = \lambda_{\pm}(k)N$, we can replace the discrete sum over $l$ by an integral. We show (see Appendix \ref{sec:exact_anyN} and \ref{sec:largeN}
 for details) that 
it converges to the following form
\begin{equation} \label{rho_k3}
\rho_k(z\sqrt{N},\theta, N) \approx \frac{2^{-k}}{\pi^{3/2} k!} \int_{a_-(k)}^{a_+(k)} dx \, e^{-x^2} \, [H_k(x)]^2
\end{equation}
where $a_{\pm}(k) = \frac{(\lambda_\pm(k)-z^2)\sqrt{N}}{z \sqrt{2}}$ and $H_k(x)$ is the $k^\text{th}$ Hermite polynomial. For fixed $z$, as $N \to \infty$, the two bounds $a_+(k) \to \infty$ and $a_-(k) \to - \infty$ iff $\sqrt{\lambda_-(k)}< z<  \sqrt{\lambda_+(k)}$. If $z$ is outside this interval, both bounds
tend to either $+ \infty$ or $- \infty$ simultaneously. In the latter cases, the integral in (\ref{rho_k3}) vanishes as $N \to \infty$. In contrast, in the former case, the integral approaches a finite value $\int_{-\infty}^{\infty} dx \, e^{-x^2} \, [H_k(x)]^2 = 2^k k! \sqrt{\pi}$. Hence we conclude that the density from the $k^{\rm th}$ band at a fixed rescaled distance $z = r/\sqrt{N}$ converges to 
\begin{equation} \label{rho_bulk}
\rho_k^{\rm bulk} (r,\theta, N) \approx  \frac{1}{\pi} {\cal I}_{\sqrt{\lambda_-(k)})<z< \sqrt{\lambda_+(k)}} \;,
\end{equation}
where the function ${\cal I}$ takes value $1$ if the inequality in the subscript is satisfied and $0$ otherwise. Thus the bulk density is flat (with value $1/\pi$) inside the $k^{\rm th}$ annulus $\sqrt{\lambda_-(k)}<z< \sqrt{\lambda_+(k)}$ [see Fig. \ref{combinedFig} b)].  We find in Fig.~\ref{combinedFig} c) an excellent agreement between the results obtained from the exact evaluation of the sum in (\ref{gen_density}) and (\ref{rho_k}) for $k^* = 1$ and the large $N$
bulk density in Eq.~(\ref{rho_bulk}). For a fixed $k^*\geq 1$, the sum in Eq. (\ref{gen_density}) gives a superposition of contributions of the type (\ref{rho_bulk}) for each $k\leq k^*$, leading to the ``wedding cake'' structure in Fig.~\ref{combinedFig}~b).

If $z$ is close to one of the two edges, say the left edge $\sqrt{\lambda_-(k)}$, we can estimate the limiting form of the edge density when $N \to \infty$ from the same Eq.~(\ref{rho_k3}). For this, we set
$ z^2 = \lambda_-(k) + \frac{\sqrt{2 \lambda_-(k)}}{\sqrt{N}} u$
where $u\sim O(1)$. In this case, the lower limit in the integral in Eq.~(\ref{rho_k3}) becomes $a_-(k) \approx -u$ (with $u$ measuring the scaled distance from the left edge), while the upper limit still approaches to $+ \infty$ as $N \to \infty$. Hence, we get,
\begin{equation} \label{rho_edge}
\rho_k^{\rm edge}(r,\theta, N) \approx f^{\rm edge}_k(u)
\end{equation}
where
$f_k^{\rm edge}(u) = \frac{2^{-k}}{\pi^{3/2} k!} \int_{-u}^{\infty} dx \, e^{-x^2} \, [H_k(x)]^2 $. Note that when $u \to \infty$, $f_k^{\rm edge}(u) \to 1/\pi$, and the edge density matches smoothly with the bulk density. In Fig. \ref{combinedFig} we have zoomed in on the left edge of $k=1$ layer and plotted the scaling function $f^{\rm edge}_1(u)$ in the inset, which clearly shows a kink where $df_1^{\rm edge}/du = 0$. For the $k^{\rm th}$ layer, setting 
$df^{\rm edge}_k(u)/du = 0$ (which implies $H_k(-u)= 0$), it follows that there will be $k$ kinks in $f^{\rm edge}_k(u)$ whose locations coincide with 
the $k$ zeros of $H_k(-u)$. The scaling function $f^{\rm edge}_k(u)$ is actually universal in the sense that it does not depend on $c$ and $M$ explicitly.  
In fact, in the special case $\gamma=0$ and $\nu=1$, but with fixed $N$ (the classical Landau problem) -- hence not in the scaling limit discussed here --, the 
edge density $\rho_k$ for the $k^{\rm th}$ Landau level was studied in \cite{dunne1994edge} and similar kinks were found for finite $N$, but the scaling function $f^{\rm edge}_k(u)$ was not computed (see also \cite{haimi2013polyanalytic} in the mathematics literature in the  context of polyanalytic Ginibre ensembles). 

Furthermore, an interesting limiting shape emerges for $f_k^{\rm edge}(u)$ in the scaling limit of $k$ large and $u$ large but with the ratio $y = u/\sqrt{2k}$ fixed. 
In this case, we find~(see Appendix \ref{sec:largeN} for details)
\begin{eqnarray} \label{largek}
\lim_{k \to \infty} f_k^{\rm edge}(\sqrt{2k}\,  y) =  \dfrac{1}{\pi^2} \cos^{-1}(-y) \;{\cal I}_{-1 < y < 1} \;,
\end{eqnarray}
and for $y>1$ it takes a value of $1/\pi$.
Interestingly, a similar shape appeared in the description of a propagating front in the quantum evolution of a spin chain (equivalent to free fermions on a lattice) \cite{antal1999transport,antal2008logarithmic,eisler2013full,hunyadi2004dynamic,mukherjee2018quantum}. It turns out that there is yet another interesting scaling regime close to the two endpoints $u \approx \pm \sqrt{2k}$. For example, setting $u=-\sqrt{2k} + \frac{w}{\sqrt{2} k^{1/6}}$ with $w = O(1)$
\begin{eqnarray} \label{Airy}
\lim_{k \to \infty} k^{1/3} \pi\, f^{\rm edge}_k\left(-\sqrt{2k} + \frac{w}{\sqrt{2} k^{1/6}}\right) = {\cal F}(w) 
\end{eqnarray}
where ${\cal F}(w) = \left( [{\rm Ai}'(-w)]^2 + w {\rm Ai}^2(-w) \right)$ where ${\rm Ai}(z)$ denotes the Airy function. Interestingly, the same
scaling function describes the tail of the density of eigenvalues (centered and scaled) in the Gaussian Unitary Ensemble of RMT \cite{bowick1991universal,forrester1993spectrum}.

\begin{figure}[t]
\includegraphics[width = 1.0\linewidth]{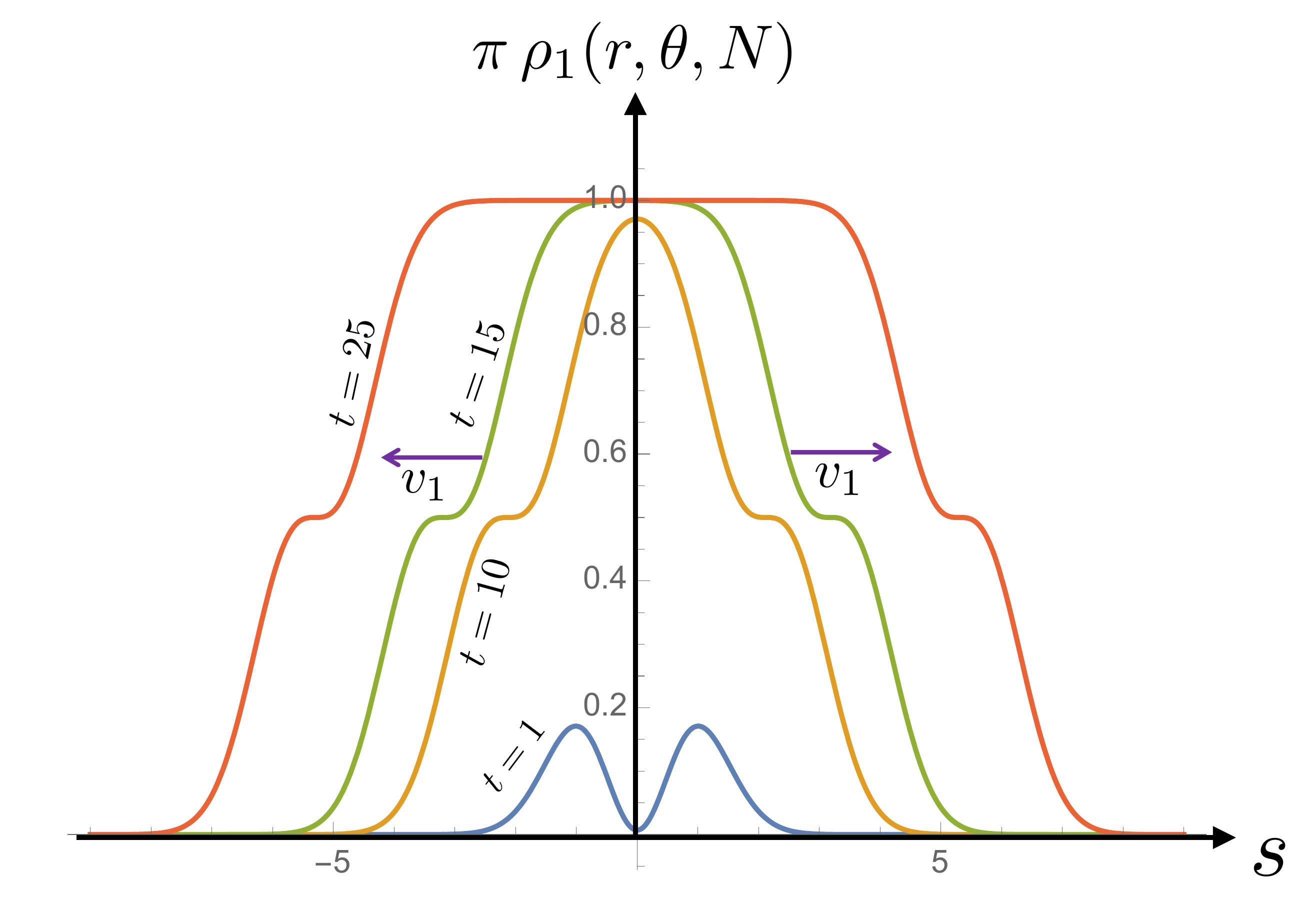}
\caption{Density profile in Eq.~\eqref{delta_rho_6} for $M=5$ plotted as a function of the scaled distance $s$ \cite{footnote_scaled_s} for increasing values of $t = 1, 10, 15$ and $25$. As $t$ increases, the scaled density approaches the constant value $1$ for $|s| < v_1 t$ and decays rapidly to $0$ for $|s| > v_1 t$. For $t \gg 1$, the forward and backward fronts separating the constant density $1/\pi$ and the zero-density outside move ballistically in opposite directions with a constant speed $v_1$.}\label{fig_wave}
\end{figure}

\section{Crossing the critical line in the $(M,c)$ plane} 
When $k^*$ changes from $k^*=0$ to $k^*=1$ (which means a new band is included below the Fermi level), one may wonder how the density profile changes from a one-layered structure to a two-layered structure. When one crosses this critical line $c=c_1(M) \equiv c_1$, the second layer appears on top of the first layer (Fig.~\ref{combinedFig}). Here, we describe the evolution of the 
density profile of this newly formed droplet as a function of the distance $c_1-c$ below the critical line $c_1$ for fixed $4<M<12$~(see Appendix \ref{sec:tw} for details).  
As $k^*$ changes from $0$ to $1$, the Fermi level $\mu$ exceeds the value $\mu = 2 + \sqrt{c_1 M}$ by a small amount $\delta$: 
$\mu = 2 + \sqrt{c_1 M} + \delta$ where $\delta \ll 1$. As $k^*$ jumps from $0$ to $1$, we find that the additional macroscopic density in the second layer  appears over the scaled region 
$\sqrt{\frac{c_1}{M}} - v_1\sqrt{\delta} < z^2 <\sqrt{ \frac{c_1}{M}} + v_1 \sqrt{\delta}$ where $v_1 = \frac{\sqrt{2}}{M} (c_1 M)^{1/4}$ and $z=r/\sqrt{N}$. Therefore, the center of the second layer appears at $z_c = (c_1/M)^{1/4}$. Here, we give a scaling description of this density in the second layer just after its appearance, i.e., in the limit $\delta \to 0$.  Let $z^2 = \sqrt{\frac{c_1}{M}} + \epsilon$, where $\epsilon$ measures the distance from the center of the second layer.  We analyse Eq. (\ref{rho_k}) with $k=1$ by replacing, for large $N$, the sum by an integral and evaluating it by the saddle point method (see Appendix \ref{sec:tw} for details). This leads to the following density profile of the droplet (for a plot see Fig. \ref{fig_wave})
\begin{eqnarray} \label{delta_rho_6}
\rho_1(z \sqrt{N},\theta,N) \approx \frac{1}{\pi} \left[ F_1(s+v_1 t )    -  F_1(s-v_1 t)   \right] \;,
\end{eqnarray}
where
$F_1(z) =  \frac{1}{2} \left[{\rm erfc}(z) - \frac{2}{\sqrt{\pi}} z \, e^{-z^2} \right]$ with ${\rm erfc}(z) $ being the complementary error function.  Here 
$s=\epsilon\sqrt{N/2} (M/c_1)^{1/4}$ is the scaled distance measured from the center of the droplet, while 
$t= \sqrt{N/2} (M/c_1)^{1/4} \sqrt{\delta}$ is proportional to $c_1-c > 0$, measuring the deviation from the critical line. If we interpret $s$ and $t$ as space and time, the density profile in Eq. (\ref{delta_rho_6}) has an interesting interpretation: the two edges of this profile move ballistically away from the droplet center with a constant speed $v_1$. At large $t$, the widths of these ``solitonic'' fronts remain of $O(1)$ while the height of the density behind the fronts approaches a constant value $1/\pi$ (see Fig. \ref{fig_wave}). This picture can be easily generalized to other critical lines in the $(M,c)$ plane (see Appendix \ref{sec:tw} for details).

\section{Conclusions}

To conclude, we have shown analytically that the average density profile in the ground-state of $N$ {\it noninteracting} fermions 
in a rotating trap exhibits a rich multi-layered ``wedding cake'' structure, as more and more Landau levels participate in the ground state 
by increasing $N$, leading to a highly interesting phase diagram in the parameter space. 
This non-trivial density profile owes its origin entirely to quantum effects, and can not be obtained from a simple Local Density/Thomas-Fermi
approximation. It would be interesting to study the effect of the inclusion of more and more Landau levels on other observables, going
beyond the one-point function studied in this paper, such as the number variance and the entanglement entropy (see for e.g., \cite{lacroix2019rotating,rodriguez2009entanglement,charles2019entanglement,leschke2020asymptotic}). We note that strongly interacting 
bosons and fermions have been studied experimentally in rotating traps leading in particular to the formation of vortex lattices \cite{rotate_cornell,rotate_zwierlein}.
In our case, there is a hole in the density at the center of the trap, but this is due to the repulsive inverse square interaction $\gamma/(2 r^2)$ and it is not
related to a vortex. It will be challenging to see how interactions can change the above scenario, in particular leading to the generation of vortices.

\acknowledgements

We thank Gautam Mandal and Takeshi Morita for very useful discussions at the initial stage of this work. MK acknowledges support from the project 6004-1 of the Indo-French Centre for the Promotion of Advanced Research (IFCPAR), 
Ramanujan Fellowship (SB/S2/RJN-114/2016), 
SERB Early Career Research Award (ECR/2018/002085) and SERB Matrics Grant (MTR/2019/001101) from the Science
and Engineering Research Board (SERB), Department of
Science and Technology, Government of India. MK thanks the hospitality of Laboratoire de Physique, Ecole Normale Sup\'erieure (Paris). This research was supported by ANR grant ANR-17-CE30-0027-01 RaMaTraF. MK acknowledges support of the Department of Atomic Energy,
Government of India, under project no. RTI4001.

\appendix

\section{Model and basic properties} 
\label{sec:mp}


As mentioned in the main text, our starting Hamiltonian is,  
\begin{equation}
\label{ham_S}
\hat H = \frac{{p}^2}{2} + \frac{1}{2} \omega^2 {r}^2 + \frac{\gamma}{2\hat{r}^2}- \Omega {L}_z
\end{equation}
where
${L}_z= x p_y- y p_x = -i  (x \partial_y - y \partial_x)$
is the $z$-component of the angular momentum, $\gamma$ characterises the repulsive-like potential at the centre (inverse-square type), 
$\omega$ is the trap frequency and $\Omega$ is frequency at which the trap rotates around the vertical axis. For convenience, we have set the mass $m=1$ and also $\hbar = 1$. Throughout the work, we will consider the case in which the inverse square central potential is very large, i.e., 
\begin{equation}
{\gamma =c N,\,\, c \sim O(1),\,\,\, \text{regime of interest}} 
\label{eq:gammabeta_S}
\end{equation}

In polar coordinates ($r,\theta$), the Hamiltonian reads,  
\begin{align}
\label{ham-pol_S}
\hat H= -\frac{1}{2}\left(\partial_r^2 + \frac{1}{\hat{r}} \partial_r \right) +
\frac{p_\theta^2}{2  \hat{r}^2} +
\frac{1}{2}  \omega^2 \hat{r}^2 + \frac{\gamma}{2\hat{r}^2} - \Omega {{L}_z}
\end{align}
where
\begin{equation}
{ L}_z= p_\theta= -i \partial_{\theta}
\end{equation}
%
The first goal is to find eigenstates and eigenvalues of Eq.~\ref{ham-pol_S}.  Let us substitute, 
\begin{equation}
\psi (r, \theta) = \psi (r) e^{il\theta}
\end{equation}
{where $l=0, \pm 1, \pm 2, ...$ are integers because the wave function needs to respect $2\pi$ periodicity in the angular direction}.
Then, we get, 

\begin{equation}
\hat{H}\psi (r) = \Bigg[ -\frac{1}{2}\left(\partial_r^2 + \frac1r \partial_r\right)  +
\frac{1}{2}  \omega^2 r^2 + \frac{\gamma+l^2}{2r^2} -l  \Omega    \Bigg] \psi (r)
\label{ham-pol1_S}
\end{equation}
Hence, the equation we need to solve is, $\hat{H}\psi (r) = E\psi (r)$ which gives us, 
\begin{equation}
\Bigg[ -\frac{1}{2}\left(\partial_r^2 + \frac1r \partial_r\right)  +
\frac{1}{2}  \omega^2 r^2 + \frac{\gamma+l^2}{2r^2} -l  \Omega    \Bigg] \psi (r)=E  \psi (r)
\label{eq:hpsi_S}
\end{equation}
To reduce the above eigenvalue equation (Eq.~\ref{eq:hpsi_S}) to a standard form, we make the following transformation
\begin{equation} \label{transfo_S}
\psi(r) = e^{-\omega r^2/2} r^{\sqrt{\gamma + l^2}} G( \omega r^2) \;.
\end{equation}
It is then easy to see that $G(z)$ satisfies the differential equation
\begin{eqnarray}\label{hypergo_S}
z G''(z) + (b-z) G'(z) - a \, G(z) = 0 \;,
\end{eqnarray}

\begin{eqnarray}\label{hypergo_S}
{\rm where} \quad\quad
\begin{cases}
&a = \frac{1}{2} \left[1 + \sqrt{\gamma+l^2} - \frac{E + \Omega \, l}{\omega}\right] \;, \\
&\\
& b= 1 + \sqrt{\gamma+l^2} \;.
\end{cases}
\end{eqnarray}

This is a standard confluent hypergeometric differential equation whose general solution is given by the linear combination of
two independent solutions as follows \cite{gradshteyn2014table}

\begin{eqnarray} \label{sol_hyper_S}
G(z) = A_1 \, z^{1-b} \, M(a-b+1,2-b,z) + A_2 \, M (a,b,z) \nonumber \\
\end{eqnarray}
where $A_1$ and $A_2$ are two arbitrary constants and 
\begin{eqnarray}\label{hypergeo_M_S}
M(a,b,z)= \sum_{p=0}^{\infty} \frac{(a)_p}{(b)_p} \frac{z^p}{p!}= 1+ \frac{a}{b} z+ \frac{a(a+1)}{b(b+1)} \frac{z^2}{2!} +...\nonumber \\
\end{eqnarray}
  is the Kummer's
confluent hypergeometric function. Here, $(a)_p, (b)_p$ are Pochhammer symbols, i.e., $(a)_p=\frac{\Gamma(a+p)}{\Gamma(a)}$ where $\Gamma(x)$ is a Gamma function. Note that the arguments of the two functions in (\ref{sol_hyper_S}) are different. The function $M(a,b,z)$ has the following asymptotic behaviors
\bea \label{M_asympt_S}
M(a,b,z) \approx
\begin{cases}
&1 + \dfrac{a}{b}z + O(z^2) \;, \; z \to 0 \\
& \\
& \dfrac{e^z \, z^{a-b}}{\Gamma(a)} \;, \;\quad\quad\quad\; z \to \infty  \;.
\end{cases}
\eea
Hence the most general solution
for the eigenfunction $\psi(r)$ in Eq. \ref{transfo_S} reads
\bea \label{gen_psi_S}
\psi(r) &=& e^{-\omega r^2/2} r^{\sqrt{\gamma + l^2}} \times \nonumber \\ && \big[ A_1 \, (\omega\, r^2)^{1-b} \, M(a-b+1,2-b,\omega\, r^2) \nonumber \\&+& A_2 \, M (a,b,\omega\, r^2)  \big] \;,
\eea
where $a$ and $b$ are given in Eq. \ref{hypergo_S}. 

To fix these unknown constants $A_1$ and $A_2$, we first consider the behavior of $\psi(r)$ as $r \to 0$. Using Eq. \ref{M_asympt_S} we see that, 	as $r \to 0$, $\psi(r) \sim A_1 \omega^{-\sqrt{\gamma+l^2}} r^{-\sqrt{\gamma+l^2}}$. However, the eigenfunction must be square-integrable, i.e., $2\pi \int_0^\infty \psi^2(r) r\,dr$ should be finite. Substituting the small $r$ behavior, we see that the integral behaves as $r^{2(1-\sqrt{\gamma+l^2})}$ in the lower limit $r \to 0$. Hence, since $l = 0, \pm 1, \cdots$, the integral is divergent for all $\gamma > 1$. And this is indeed the case in our problem where $\gamma$ is scaled as $\gamma = c N$ where $c = O(1)$ and $N \to \infty$ (see Eq. \ref{eq:gammabeta_S}). Hence we must have $A_1=0$. Therefore the solution now reads
\bea \label{gen_psi_2_S}
\psi(r) = A_2 \, e^{-\omega r^2/2} r^{\sqrt{\gamma + l^2}}  \, M (a,b,\omega\, r^2)   \;.
\eea
We now consider the other limit $r \to \infty$. Substituting the asymptotic behaviour given in Eq. \ref{M_asympt_S} in Eq. \ref{gen_psi_2_S}, we find that 
\begin{eqnarray} \label{psi_large_r_S}
\psi(r) \approx \frac{A_2}{\Gamma(a)} \omega^{a-b} r^{2a-b-1} e^{\omega r^2/2} \;.
\end{eqnarray}
Clearly, the integral $2 \pi \int_0^\infty \psi^2(r) r \, dr$ diverges at the upper limit $r \to \infty$, provided $\Gamma(a)$ is finite. Hence, to cure this
divergence, we must choose $|\Gamma(a)| = +\infty$, which means that $a = -k$ where $k = 0, 1, 2, \cdots$ is a non-negative integer. In fact, this is the quantisation condition. In fact, when $a=-k$ the function $M(a = -k,b,z)$ is a polynomial of degree $k$ and the wave function is square integrable. The quantisation condition $a=-k$, using Eq. \ref{hypergo_S}, reads

\begin{equation}
\frac{\sqrt{\gamma +l^2}}{2}-\frac{l \Omega }{2 \omega }-\frac{E }{2 \omega }+\frac{1}{2}= -k  \;. 
\label{eq:quant_S}
\end{equation}
The normalization condition fixes the constant $A_2 = c_{k,l}$, which depends on both quantum numbers $k$ and $l$. Hence, summarising, 
the complete set of eigenfunctions are given by 
%
%
%
\begin{equation}
\psi_{k,l}(r,\theta)= c_{k,l} r^{\lambda} e^{-\omega r^2/2} 
M(-k, 1+\lambda, \omega r^2) e^{i l \theta},
\label{eq:Mfunc_S}
\end{equation}
where $ \lambda = \sqrt{\gamma+l^2} $ with the associated eigenvalues from Eq.~\ref{eq:quant_S} 
\begin{equation}
E_{k,l}= \omega[2k+1+\sqrt{\gamma+l^2}]- \Omega l \;.
\label{eq:eigenvaluesg_S}
\end{equation}

Without loss of generality, we will set $\omega=1$ (i.e., the energies are expressed in units of $\omega$) and introduce $\nu \equiv \Omega / \omega  < 1$. Note also that Kummer's
confluent hypergeometric function are related to  generalized Laguerre polynomials as,
\begin{equation}
M(-k, 1+\lambda, r^2) = \frac{\Gamma(k+1) \Gamma(1+\lambda)}{\Gamma(1+k+\lambda)}  L_k^{\lambda}(r^2) \;.
\end{equation}
{Therefore, expressing Eq.~\ref{eq:Mfunc_S} in terms of generalized Laguerre polynomials is preferable since these functions have an orthonormality condition that turns out to be useful
\begin{equation}
\int_{0}^{\infty} dx\, x^{\lambda} e^{-x}  L_k^{\lambda}(x)  L_{k'}^{\lambda}(x)dx = \frac{\Gamma(k+\lambda+1)}{\Gamma(k+1)}\delta_{kk'} \;.
\end{equation}}
The normalisation requirement $ 2\pi \int_0^{\infty} r dr |\psi_{k,l}(r) |^2=1$ finally gives, 
\begin{equation}
{\psi_{k,l}(r,\theta) = a_{k,l} L_k^{\lambda}(r^2) r^\lambda e^{-r^2/2} e^{i l \theta}}
\label{eq:meigf_S}
\end{equation}
with
\begin{equation}
a_{k,l}^2 = \frac{\Gamma(k+1)}{\pi \Gamma(k+1 + \lambda)} 
\label{eq:meigf_S_e}
\end{equation}

The associated eigenvalues are now expressed as,
\begin{equation}
{E_{k,l}= 2k+1+\sqrt{\gamma+l^2}- \nu l}
\label{eq:meigenvaluesg_S}
\end{equation}
Eq.~\ref{eq:meigf_S} and Eq.~\ref{eq:meigenvaluesg_S} form the complete solution of our system.  In what follows, we will analyse the energy levels (Eq.~\ref{eq:meigenvaluesg_S}) of the system.

\section{Analysis of energy levels and the ground state} 
\label{sec:elev}

\begin{figure*}[t]
\centering
    \includegraphics[width=.5\linewidth]{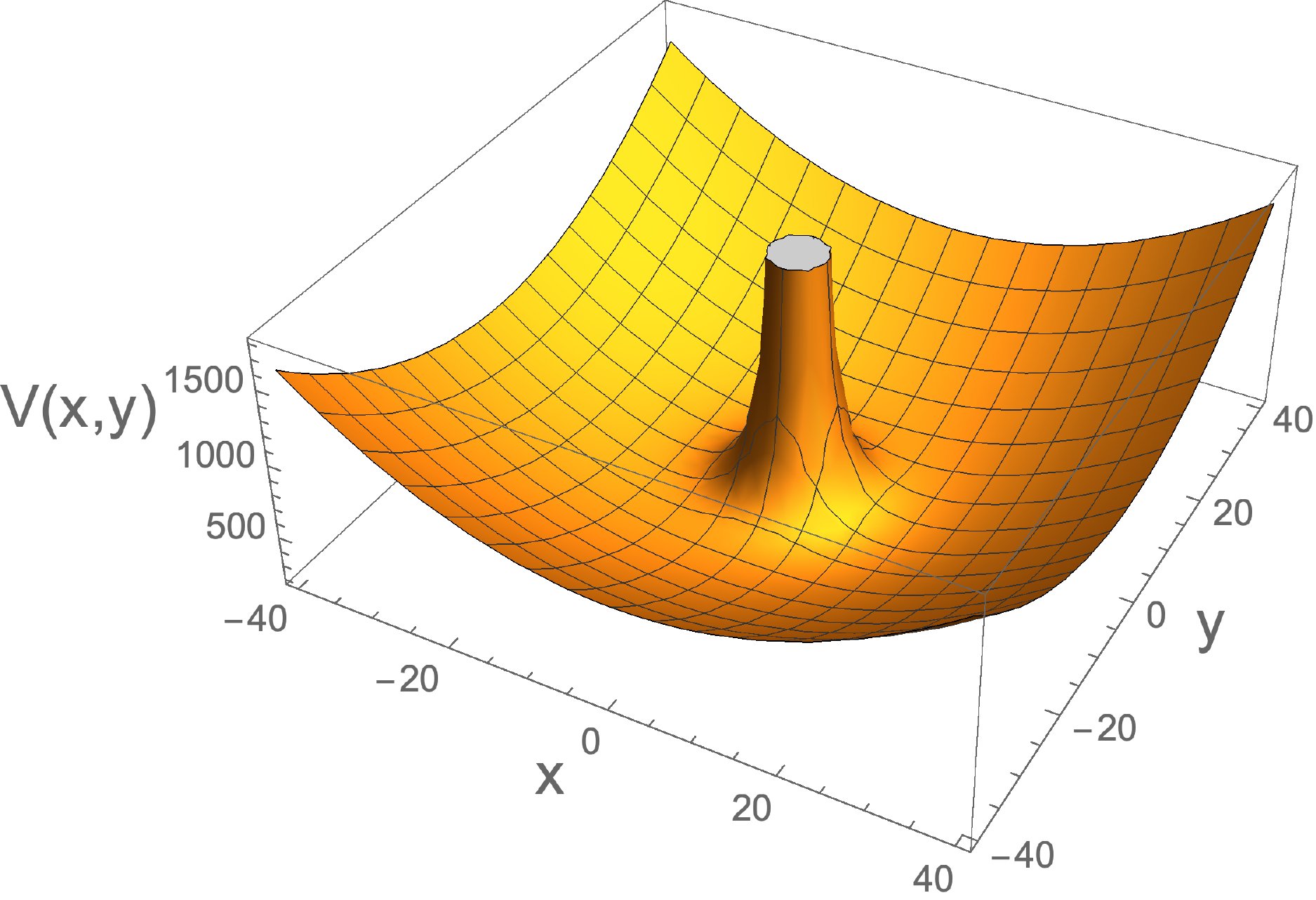}\quad\quad
    \includegraphics[width=.35\linewidth]{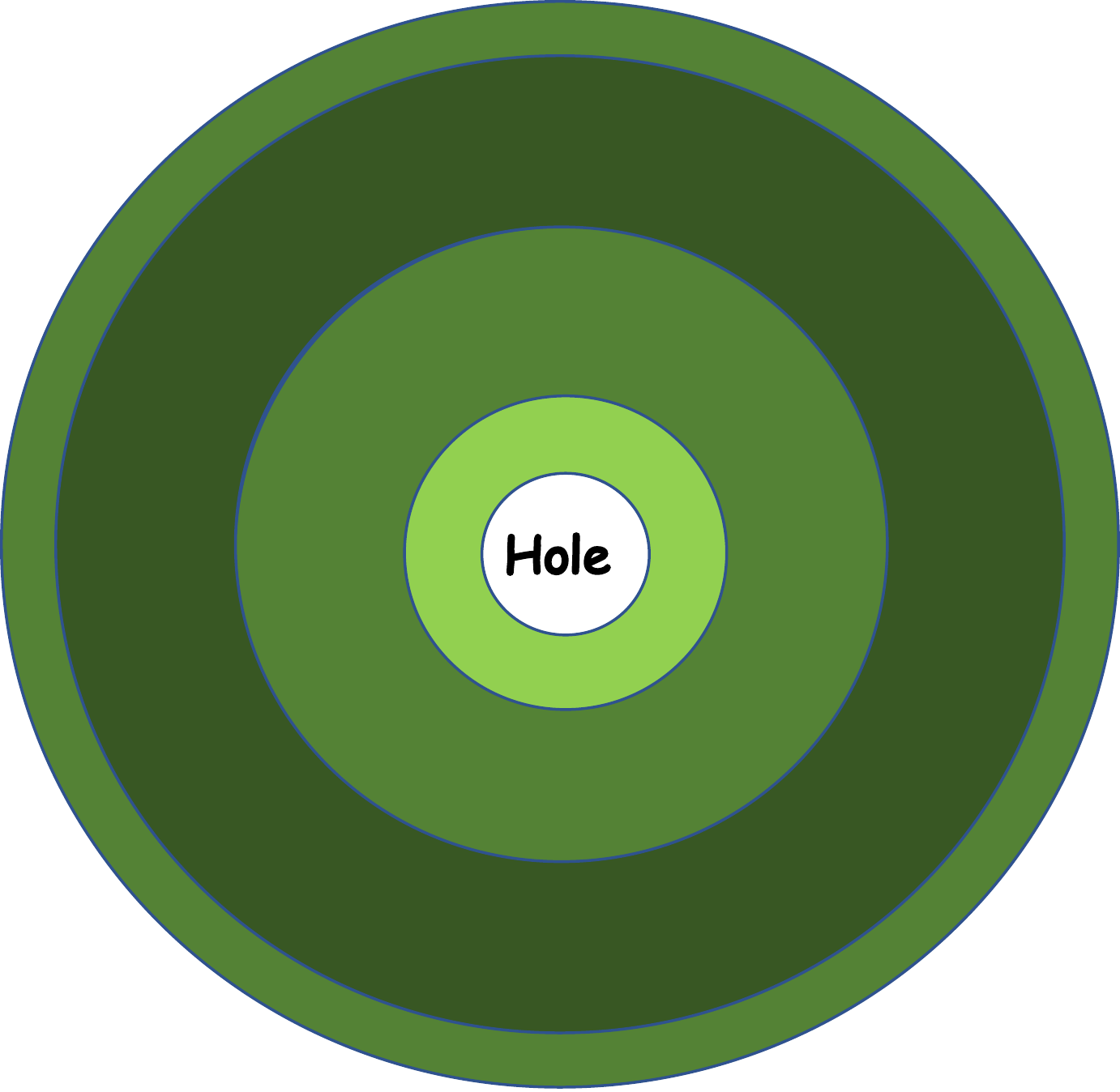}
    \caption{(Left) The external potential $V(r) =  \frac{1}{2} \omega^2 r^2 + \frac{\gamma}{2r^2}$ from Eq.~\ref{ham_S} is plotted for visualization purposes. We took $c=100$ and $N=400$. We see the highly repulsive central potential that eventually causes a hole/empty region. (Right) Here, we show a schematic figure (top view) showing the fermions in 2D. The formation of the central hole and multiple layers and multiple edges is the key finding and property of the underlying Hamiltonian (Eq.~\ref{ham_S}). This is certainly missed via a traditional Local Density Approximation (see also Fig.~\ref{figcomp_S}). }
    \label{fig:origrhosumschem_S}
\end{figure*}

\begin{figure*}[t]
\centering
    \includegraphics[width=.45\linewidth]{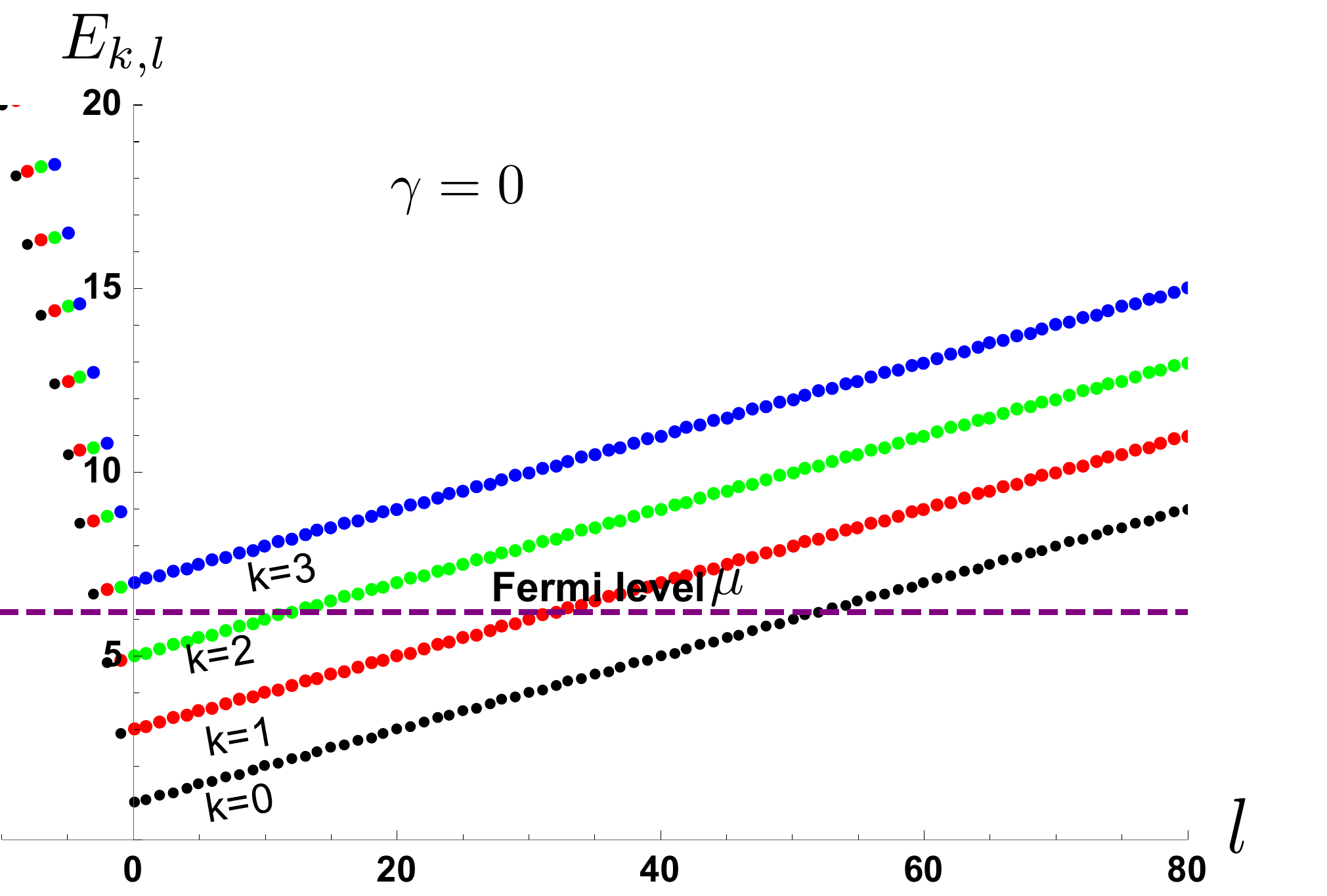}
    \includegraphics[width=.45\linewidth]{fig0_disp_nonzero_supp.pdf}
    \caption{(Left) Energy levels for the case $\gamma = 0$, $\nu=0.9$ and $\mu = 6.2$ (Eq.~\ref{eq:meigenvaluesg_S}). (Right) Energy levels for the case $\gamma = 5$, $\nu=0.9$ and $\mu=7.5$ (Eq.~\ref{eq:meigenvaluesg_S}).  For both figures, the purple (dashed) line shows the Fermi level upto which we are allowed to fill fermions. The black, red, green and blue curves represent $k=0,1,2,3$ bands respectively. The figures shows that if one fixes the Fermi level appropriately, then only three energy bands ($k=0,1,2$) play a role.  Instead of fixing the Fermi level, one can alternatively fix the number of Fermions $N$.  These figures demonstrate the dramatic difference between $\gamma=0$ and $\gamma \neq 0$ case. }
    \label{fig:disp_S}
\end{figure*}

In this section, we will analyse the energy levels (Eq.~\ref{eq:eigenvaluesg_S}) and discuss the ground state for a system which has $N$ fermions. To start with, let us recap the $\gamma = 0$ case.

\subsection{$\gamma =0$ case}

For $\gamma = 0$, Eq.~\ref{ham_S} reduces to the Hamiltonian considered in Refs.~\cite{ho2000rapidly,aftalion2005vortex,lacroix2019rotating}. The corresponding eigenfunctions were computed not in the polar coordinates, rather in the coordinates $(z,  \bar{z})$ where $z = x + i y$. In this representation, the eigenfunctions read (see e.g. Refs.~\cite{ho2000rapidly,aftalion2005vortex,rodriguez2009entanglement,lacroix2019rotating})
\begin{eqnarray} \label{zzbar_S}
\psi_{n_1, n_2}(z, \bar z) = A_{n_1, n_2} e^{z \bar z/2} \partial^{n_1}_{\bar z} \partial^{n_2}_z e^{- z \bar z} \;,
\end{eqnarray}
with the associated eigenvalues (in units such that $\omega=1$)
\begin{equation}
E_{n_1,n_2} = 1+ (1-\nu)n_1 +  (1+ \nu)n_2 \;.
\label{energyn1n2_S}
\end{equation}
where $n_1 = 0, 1, 2, \cdots$ and similarly $n_2 = 0, 1, 2, \cdots$ where $0< \nu = \Omega/\omega \leq 1$. This last condition follows
from the fact that for $\nu > 1$, the system is ``unstable'' in the sense that the fermions can ``fly away". 
Another important point one can observe is that if $\omega = \Omega$, then it becomes the Landau problem (free electrons in perpendicular magnetic field) with energy levels given by, 
\begin{equation}
E_{n_1,n_2}^{\text{Landau}} = 1+  2\Omega n_2,\,\, \text{Landau problem}
\end{equation}

The lowest Landau level (LLL) is given by $n_2=0$. For a given $n_2$, there is a $N-$fold degeneracy. We do not want degeneracy and therefore it can be lifted by choosing $\Omega<\omega$. 
This problem can also be alternatively solved in the polar coordinates discussed in the previous section. To see the connection between these two
representations, we put $\gamma=0$ in Eq.~\ref{eq:eigenvaluesg_S} and get 
\begin{equation}
E_{k,l}=(2k+1+ |l|)- \nu l
\label{energykl_S}
\end{equation}
where, $k=0,1,2...$ and $l=0,\pm1,\pm 2...$. Comparing Eq.~\ref{energykl_S} and Eq.~\ref{energyn1n2_S}, we get, 
\begin{equation}
n_1-n_2 = l,\,\quad\quad n_1+n_2 = 2k+|l|
\end{equation}
which implies,
\begin{equation}
n_1 =k+\frac{l+|l|}{2},\,\quad \quad n_2 = k+\frac{|l|-l}{2} \;.
\label{eq:nk_tran_S}
\end{equation}
Therefore the LLL $n_2 = 0$ and $n_1=0,1,2...$ corresponds to $k=0$, $l=0,1,2...$
Note that, when $k=0$, we have two branches (positive and negative $l$), 
\begin{eqnarray}
E_{k=0,l} =  (1+|l|)-\nu l
\end{eqnarray}
thus the LLL ($n_2 = 0$ and $n_1=0, 1, 2, \cdots$) corresponds to the right branch ($l \geq 0$) and $k=0$ in the polar representation of the eigenfunctions. 
The left panel of Fig.~\ref{fig:disp_S} shows the energy levels for the case of $\gamma =0$.

\subsection{$\gamma \neq 0$ case assuming $\gamma = c N$ where $c \sim O(1)$}

Now, we discuss the case with $\gamma \neq 0$. This case turns out to be quite non-trivial. We recap that the energy levels are given by (Eq.~\ref{eq:eigenvaluesg_S}), 

\begin{equation}
E_{k,l}= 2k+1+\sqrt{\gamma+l^2}- \nu l
\label{eq:dispgammasup_S}
\end{equation}
For a given $k$, $E_{k,l}$ (Eq.~\ref{eq:dispgammasup_S}, with energy shifted by 1 for convinience) has a minimum at $l=l^*$ where,
\begin{equation}
l^* = \frac{\nu}{\sqrt{1-\nu^2}} \sqrt{\gamma}
\end{equation}

Note that $l^*$ is independent of $k$ and the energy of the $k^{\text{th}}$ band at this minimum is given by,
\begin{equation} \label{E_lstar_S}
E_{k, l^*} = 2 k + \sqrt{(1-\nu^2)\gamma} \;.
\end{equation}
We fix the Fermi level at $\mu$. By varying $\mu$, we can intersect the energy spectrum $E_{k,l}$ at different points. As $\mu$ increases, more and more $k$-bands of the spectrum become lower than the Fermi level and hence should be included in the construction of the many-body ground-state. The right panel of Fig.~\ref{fig:disp_S} shows the energy levels for the case of $\gamma  \neq 0$. It intersects the $k^{\text{th}}$ band at two points $l_{\pm}(k)$ along the $l$-axis which can be easily computed by setting $E_{k,l} = \mu$ and we get,
\begin{equation} \label{lpmk_1_S}
l_{\pm}(k) = \frac{\nu(\mu - 2k) \pm \sqrt{(\mu-2k)^2 - \gamma(1-\nu^2)}}{1-\nu^2} \;.
\end{equation}
Note that if the Fermi surface has to intersect at least one band, we must have $E_{0,l^\star}<\mu$ which indicates that,
\begin{equation} \label{mubigger_S}
\mu > \sqrt{(1-\nu^2) \gamma} \;. 
\end{equation}
For a fixed $\mu$, the number of bands $k^*+1$ below $\mu$ can be obtained by setting, 
\begin{eqnarray} \label{kc_S}
E_{k^*,l^\star} &=& 2k^* + \sqrt{(1-\nu^2)\gamma}  < \mu  \nonumber \\
E_{k^*+1,l^\star} &=& 2k^* + 2 + \sqrt{(1-\nu^2)\gamma}  > \mu 
\end{eqnarray}
Hence $k^*$ is given by, 
\begin{equation} \label{kc_2_S}
k^* = \rm{Int}\left[ {\frac {\mu - \sqrt{(1-\nu^2)\gamma}}{2}} \right] \;,
\end{equation}
where ${\rm Int}(x)$ denotes the integer part of $x$. Finally, the relation between the Fermi energy $\mu$ and $N$ can be obtained by counting the total number of single particle levels with energy below $\mu$. This gives,
\begin{equation} \label{rel_mu_N_S}
\sum_{k=0}^{k^*}\left[ l_+(k) - l_-(k) \right] = N  \;.
\end{equation}
Using Eq.~\ref{lpmk_1_S}, this gives,
\begin{equation} \label{rel_mu_N1_S}
\frac{2}{1-\nu^2} \sum_{k=0}^{k^*} \sqrt{(\mu - 2k)^2 - \gamma(1-\nu^2)} = N \;.
\end{equation}

It is important to note that all the above results until now are valid for arbitrary $N$, arbitrary parameters $\nu \in [0,1]$ and $\gamma>0$.

We will now work in the large $N$ limit, and set,
\begin{equation} \label{scaling_S}
{\gamma = c N \;, \; (1-\nu^2) N = M ,\,\,\, \text{Large-$N$ limit}}
\end{equation}
Note that we are taking the limit $\nu \to 1$, $N \to \infty$, keeping $(1-\nu^2) N = M$ fixed. Therefore we have just two parameters $c$ and $M$ left and we want to calculate the average density in the ground state in the limit of large $N$, for fixed $c$ and $M$. We will see that in the $(M,c)$ plane, there is a series of critical lines separating phases with different density profiles. In terms of $c$ and $M$ we thus have
\begin{equation} \label{lpmk_S}
l_{\pm}(k) = \lambda_{\pm}(k) N \;, \;\; 
\end{equation}
where
\begin{eqnarray} \label{lpmk_S_e}
\lambda_{\pm}(k) &=& \frac{(\mu - 2 k) \pm \sqrt{(\mu-2k)^2 - c M}}{M} \;,\nonumber \\
k^* &=& {\rm Int}\left[ \frac{\mu - \sqrt{cM}}{2}\right] \;.
\end{eqnarray}

Similarly the relation between $\mu$ and $N$ in Eq.~\ref{rel_mu_N1_S} becomes
\begin{equation} \label{rel_mu_N2_S}
\frac{2}{M} \sum_{k=0}^{k^*} \sqrt{(\mu - 2k)^2 - c M} = 1 \;.
\end{equation}
{For fixed $c$ and $M$, we have $\mu \sim O(1)$.  Note that if we need $ \lambda_{\pm}(k)$ in Eq.~\ref{lpmk_S} to be $O(1)$, then we had to choose the scaling $\gamma = c N$. This justifies a posteriori the scaling $\gamma = c N$ for large $N$ used in Eq.~\ref{scaling_S}. Next, we will discuss this $(M,c)$ plane and critical lines in this plane. }

\section{$(M,c)$ plane and critical lines}
\label{sec:cmp}

\begin{figure}[t]
\centering
       \includegraphics[width=.48\textwidth]{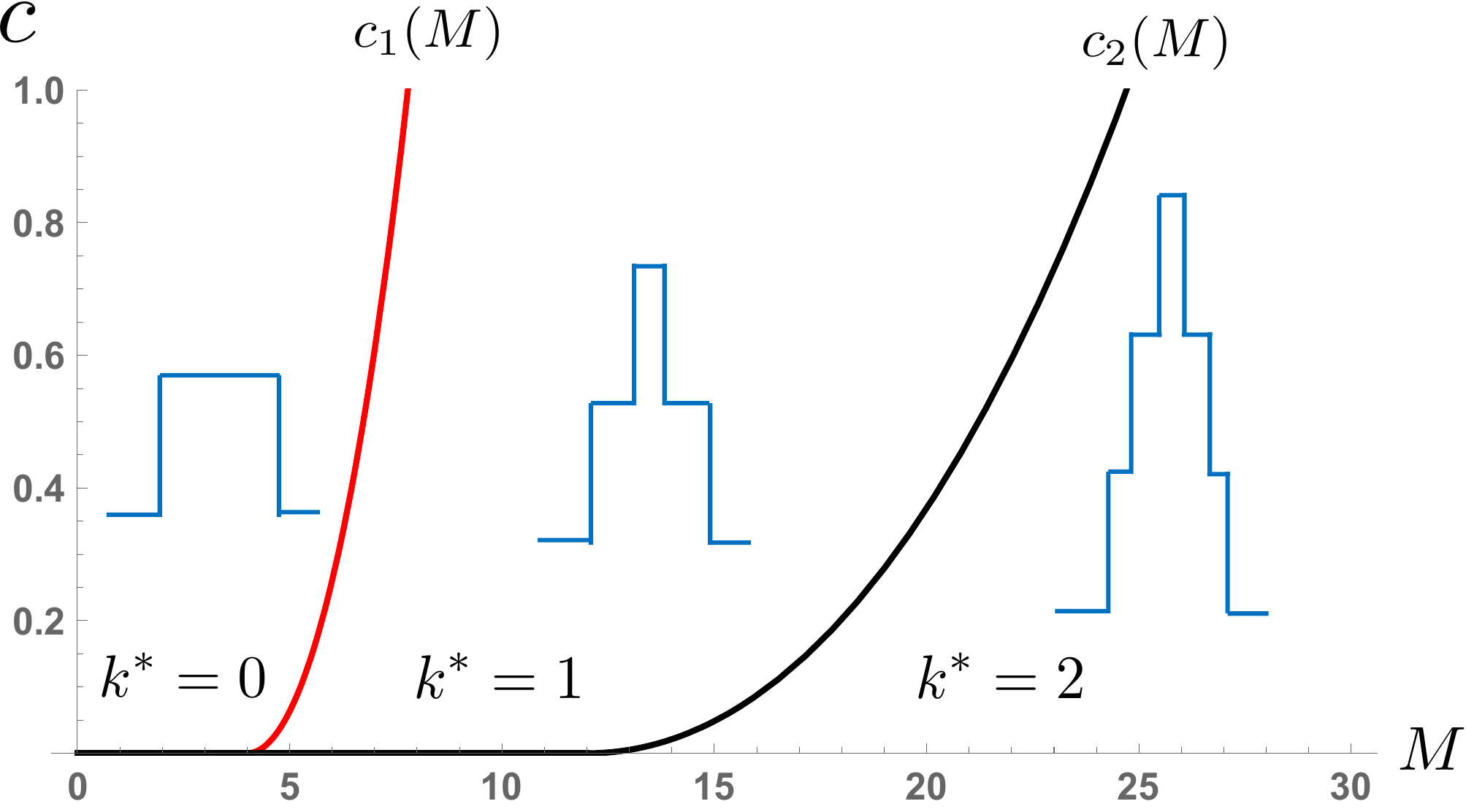}
    \caption{Phase diagram in the $(M,c)$ plane. It is divided into regions labeled by $k^* =0, 1, 2...$ denoting the number of bands ($n$) that are below the Fermi level. The lines $c_{n}(M)$ separates the regions between $k^*=n-1$ and $k^*=n$. In each of the regions, a typical (representative) density profile is shown (blue). We see that every new band creates a new layer in the density.}
    \label{fig:cmp_S}
\end{figure}

The $(M,c)$ plane gets divided in different regions, each labeled by $k^* = 0, 1, 2, \cdots$. For example, if $k^*=0$ (only the first band $k=0$ is included in the ground state), we must have
\bea \label{k0_only_S}
\sqrt{cM} < \mu < 2 + \sqrt{cM} \;.
\eea 
The upper inequality gets violated when $\mu = 2 + \sqrt{cM}$. Substituting this value of $\mu$ in Eq.~\ref{rel_mu_N2_S} with $k^* = 0$, we get the first critical line $c=c_1(M)$ in the $(M,c)$ plane [see Fig.~\ref{fig:cmp_S}],  
\bea \label{c1M_S}
c_1(M) = 
\begin{cases}
& \dfrac{1}{M} \left( \dfrac{M^2}{16}-1\right)^2 \;\; , \;\; \;\; M \geq 4 \; \\
& \\
& 0 \;\;\; \hspace*{2.5cm}M < 4 \;.
\end{cases}
\eea 
Hence if $c \geq c_1(M)$, the ground state contains only the $k=0$ band (i.e. $k^*=0$). 

Next, let us consider the case $k^* = 1$, i.e. two bands $k=0$ and $k=1$ are below the Fermi level $\mu$.
From the equation for $k^*$ in Eq.~\ref{lpmk_S} we see that $k^*={\rm Int}(\frac{\mu - \sqrt{cM}}{2})=1$ implies  
\bea \label{kc_1_S}
2 + \sqrt{cM} < \mu < 4 + \sqrt{cM} \;.
\eea
The lower limit corresponds to the critical line $c=c_1(M)$ discussed before. The upper limit gives a new critical line $c=c_2(M)$ obtained by substituting $\mu = 4 + \sqrt{c M}$ in Eq.~\ref{rel_mu_N2_S} with $k^*=1$, i.e.,
\bea \label{rel_N_mu_kc1_S}
\frac{2}{M} \left[ \sqrt{(4+\sqrt{cM})^2 - cM} + \sqrt{(2 + \sqrt{cM})^2 - cM}\right] = 1 \nonumber \\ \;.
\eea
Solving for $c = c_2(M)$ we get,
\begin{widetext}
\bea \label{c2M_S}
c_2(M) =
\begin{cases}
& \dfrac{17 M^5-416 M^3-12 \sqrt{2} \sqrt{M^{10}-48 M^8+768 M^6-4096 M^4}+2304 M}{256 M^2} \;, \; M > 12 \\
& \\
& 0 \;, \; \hspace*{10.5cm} 4< M<12
\end{cases}
\eea
\end{widetext}
This second critical line (Eq.~\ref{c2M_S}) is also plotted in Fig.~\ref{fig:cmp_S}.  
For higher values of $k^*$, one can obtain a similar formula for the critical line. For general $k^*$, the condition $k^* = {\rm Int}( \frac{\mu - \sqrt{cM}}{2})$ indicates that,
\bea \label{genkstar_S}
2 k^* + \sqrt{cM} < \mu < 2 (k^*+1) + \sqrt{c M} \;.
\eea
Setting $\mu = 2 (k^*+1) + \sqrt{c M}$ in Eq.~\ref{rel_mu_N2_S} and simplifying, one gets,
\bea \label{critical_kstar_S}
\frac{4}{M}\sum_{q=1}^{k^*+1} \sqrt{q(q+\sqrt{cM})} = 1 \;.
\eea
Solving this equation for $c$ as a function of $M$ gives the critical line $c_{k^*+1}(M)$. For $k^*=0$ and $k^*=1$ the explicit solutions are given respectively in Eqs. \ref{c1M_S} and \ref{c2M_S}. However, one can easily work out the asymptotics. For example the line $c_{k^*+1}(M)$ starts from $M^* = 2(k^*+1)(k^*+2)$. For $k^*=0$ and $k^*=1$, this gives $M^* = 4$ and $M^*=12$ respectively. For large $M$, it is easy to show from Eq.~\ref{critical_kstar_S} that,
\bea \label{largeM_S}
c_{k^*+1}(M) \approx \frac{M^3}{2^8\left[\sum_{q=1}^{k^*+1} \sqrt{q}\right]^4} \;.
\eea
Thus in the $(M,c)$ plane, we get different regions labeled by $k^* = 0, 1, \cdots$. The region between $c_{n}(M)$ and $c_{n+1}(M)$ corresponds to the region with $k^* = n$, i.e., the Fermi level has exactly $n+1$ bands below it. 
In Fig.~\ref{fig:cmp_S}, we show the $(M,c)$ plane and some critical lines that demarcates various regions. In each region, we have also sketched a typical/representative density profile. Next, we will discuss the density in the ground state.

\section{Density as a function of space (exact expression for finite $N$)}
\label{sec:exact_anyN}

We recap (Eq.~\ref{eq:meigf_S}) that the single particle wave functions can be written as

\begin{equation}
\psi_{k,l}(r,\theta) = a_{k,l} L_k^{\lambda}(r^2) r^\lambda e^{-r^2/2} e^{i l \theta} \;, 
\label{eq:meigf1_S}
\end{equation}
with
\begin{equation}
a_{k,l}^2 = \frac{\Gamma(k+1)}{\pi \Gamma(k+1 + \lambda)} \;\, \text{and} \;\, \lambda = \sqrt{\gamma + l^2}\;.
\label{eq:meigf1_S_e}
\end{equation}
%
$L_{k}^{\alpha}(x)$ are the generalized Laguerre polynomials. 
The average density in the ground state is given by the general formula,
\begin{eqnarray} \label{gen_density_S}
\rho(r,\theta,N) &=&  \sum_{k,l} |\psi_{k,l}(r,\theta)|^2  \nonumber \\ 
&=&\frac{e^{-r^2}}{ \pi} \sum_{k=0}^{k^*} \sum_{l = l_-(k)}^{l_+(k)} \frac{\Gamma(k+1)\,[L_k^\lambda(r^2)]^2 \; r^{2 \lambda}}{\Gamma(\lambda+k+1)} \nonumber \\ &=& \sum_{k=0}^{k^*} \rho_k(r,\theta,N)
\end{eqnarray}
where $l_{\pm}(k)$ are given in Eq.~\ref{lpmk_S} with $\mu$ determined from Eq.~\ref{rel_mu_N2_S}. The contribution to the density from the $k^{th}$ band is given by,
\bea \label{rho_k_S}
\rho_k(r,\theta,N) = \frac{\Gamma(k+1)\,e^{-r^2}}{\pi } \sum_{l = l_-(k)}^{l_+(k)} \frac{[L_k^\lambda(r^2)]^2 \; r^{2 \lambda}}{\Gamma(\lambda+k+1)} \,\,\,
\eea

\begin{figure*}[t]
\centering
    \includegraphics[width=.45\linewidth]{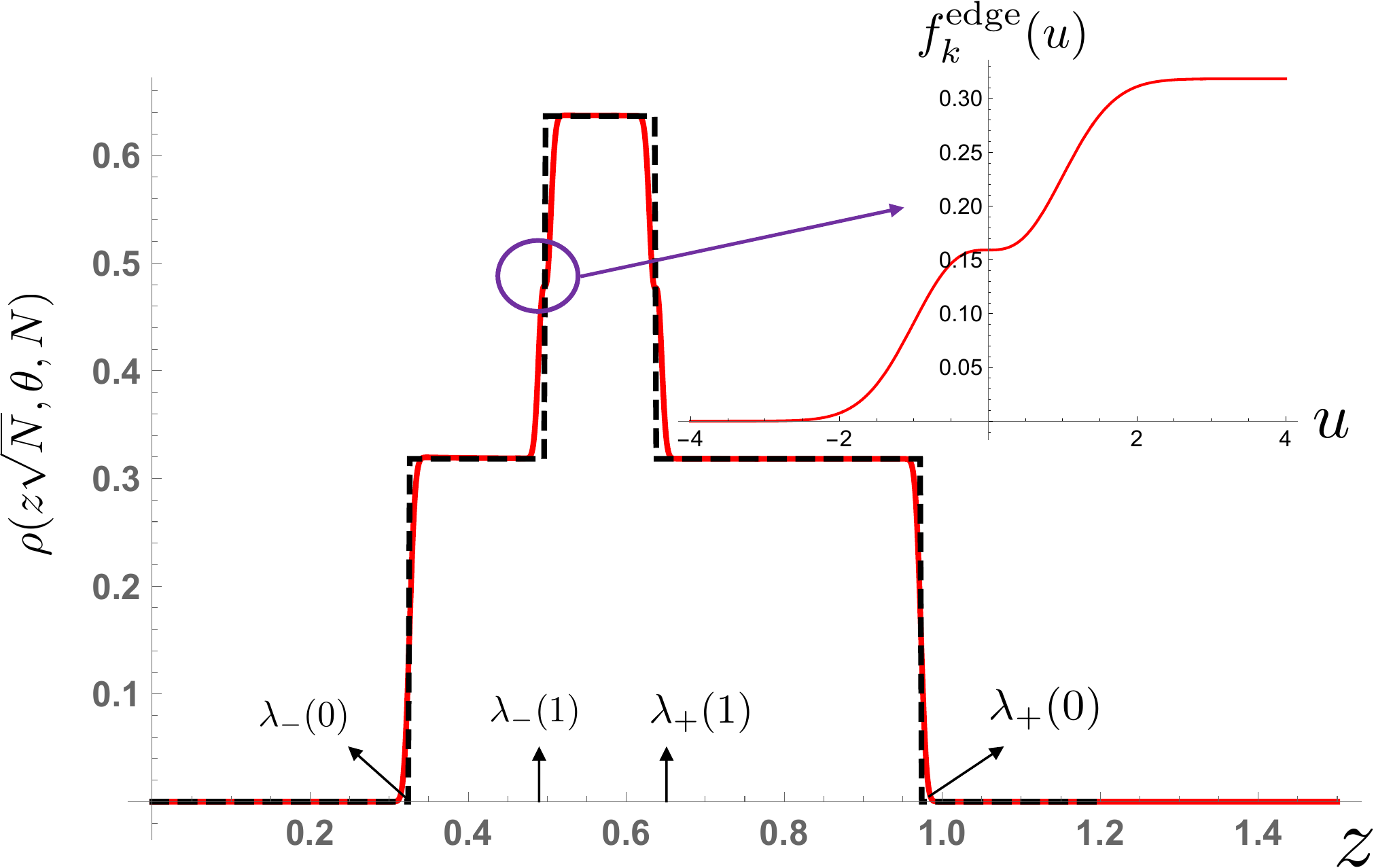}
     \includegraphics[width=.5\linewidth]{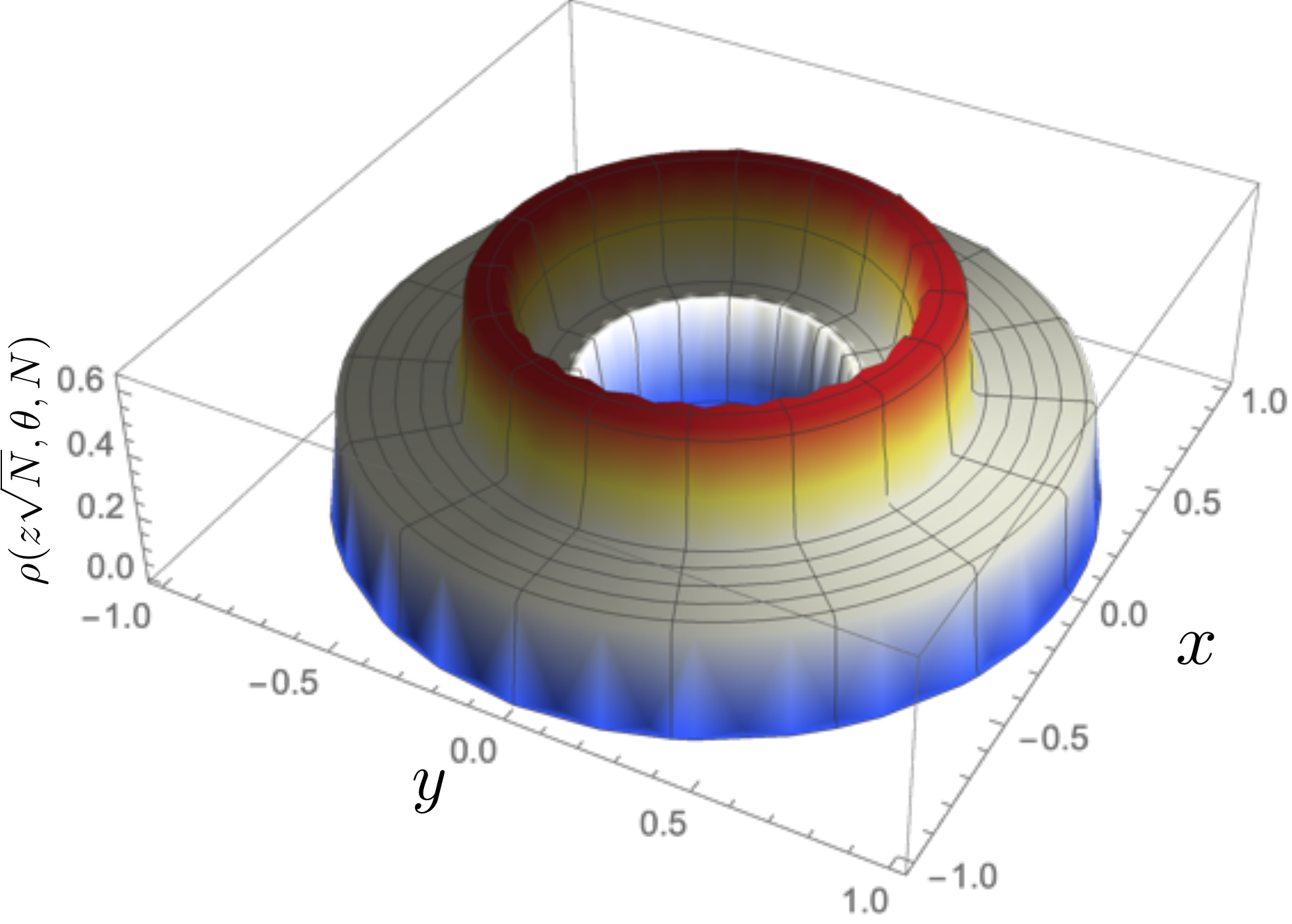}
    \caption{(Left) Plot showing the comparison between exact expression for density (Eq.~\ref{gen_density_S}, red solid) and the bulk density expression at large-$N$ (Eq.~\ref{rho_bullk_S}, black dashed). We chose, $c=1$, $M=10$ and $N=8000$ and we are in the $k^*=1$ region of Fig.~\ref{fig:cmp_S}. We also notice the two kinks (red solid) which stems from the zeros of Hermite Polynomial of degree $k=1$ in our case.  We have zoomed the location of the kink for the left edge and shown $f^{\rm edge}_1(u)$. (Right) A three-dimensional representation of exact expression for density (Eq.~\ref{gen_density_S}). We can see the non-trivial layered structure ($z=\sqrt{x^2+y^2}$).}
    \label{figcomp_S}
\end{figure*}

The above expression for density (Eq.~\ref{gen_density_S} and Eq.~\ref{rho_k_S}) is valid for any $N$ (see Fig.~\ref{figcomp_S}). In what follows, we will take the large-$N$ limit and provide further analytical insight into the form of the density.

\section{Density as a function of space in the large $N$ limit}

\label{sec:largeN}

In the large $N$ limit, noting that both $l_{\pm}(k)$ scale as $N$, we set {$l = N y$} and replace the discrete sum over $l$ by an {integral over $y$.} Furthermore, we scale $r = z \sqrt{N}$. With this change of variable, we want to first express the integrand as a function of {$y$} for fixed $z$ in the limit of large $N$. Let us start with the quantity $\lambda = \sqrt{\gamma + l^2}$. Recollecting that $\gamma = c N$ and setting {$l = N y$}, we get for large $N$,
{\bea \label{lambda_S}
\lambda \simeq N y + \frac{c}{2y} \;
\eea}
Approximating the Gamma function in Eq.~\ref{rho_k_S} by the Stirling formula {$\Gamma(z+1)\sim \sqrt{2\pi}\, e^{\big(z+\frac{1}{2}\big) \log (z)-z}$ for large $z$} and setting $r = z \sqrt{N}$, we find to leading order for large $N$,
\begin{widetext}
{\bea \label{rho_k2_S}
\rho_k(r,\theta,N) \approx \frac{\sqrt{N}\Gamma(k+1)}{\pi \sqrt{2 \pi}} \int_{\lambda_-(k)}^{\lambda_+(k)} \frac{dy}{\sqrt{y}} e^{N \left[ y \ln (z^2/y) + y -z^2\right]} (N y)^{-k} [L_k^{N y}(z^2 N)]^2 \;,
\eea}
\end{widetext}
where $\lambda_{\pm}(k)$ has been defined in Eq. \ref{lpmk_S}. In the large $N$ limit, the {integral over $y$ is dominated by a saddle point at $y=z^2$.} Therefore it is natural to make the change of variable,
{\bea \label{change_S}
y=z^2 + \sqrt{\frac{2}{N}} \, x\, z \;.
\eea}
{Therefore $Nz^2 \approx N y - x \sqrt{2 N y}$. }We can now use the following remarkable limiting formula for the generalized Laguerre polynomials,
{\bea \label{hermite_S}
\lim_{\lambda \to \infty} \lambda^{-k/2} L_k^\lambda(\lambda - \sqrt{2 \lambda} x) = \frac{2^{-k/2}}{\Gamma(k+1)} H_k(x) \;,
\eea}
where {$H_k(x)$} is the Hermite polynomial of index $k$. {Substituting $\lambda \approx N y$ and using $Nz^2 \approx N y - x \sqrt{2 N y}$ we find, using Eq.~\ref{hermite_S}, that, 
\bea \label{limit_forme_S}
\lim_{N \to \infty} (N\,y)^{-k} [L_k^{N y}(N y - x \sqrt{2 N y})]^2   = \frac{2^{-k}}{[\Gamma(k+1)]^2} H_k^2(x) \;.\nonumber \\
\eea }
Thus the integral in Eq.~\ref{rho_k2_S} reads
\begin{equation} \label{rho_k3_S}
\rho_k(r,\theta,N) \approx \frac{2^{-k}}{\pi^{3/2} \Gamma(k+1) } \int_{a_-(k)}^{a_+(k)} dx \, e^{-x^2} \, [H_k(x)]^2 \;, 
\end{equation}
where
\begin{equation} \label{rho_k3_S_e}
a_{\pm}(k) = \frac{(\lambda_\pm(k)-z^2)\sqrt{N}}{z \sqrt{2}} \;. 
\end{equation}
Therefore, the density in the $k^{\text{th}}$ band is supported on the interval $\sqrt{\lambda_-(k)}<z< \sqrt{\lambda_+(k)}$. It turns out that this expression for the density has very interesting bulk and edge properties. In the subsequent subsections we analyse these properties.

\subsection{Bulk}
\label{ssec:bulk}

 If $z$ is in the bulk, i.e. far away from these two edges, then in the large $N$ limit, the two limits $a_{\pm}(k) \to \pm \infty$. Hence the integral becomes simply $\int_{-\infty}^{\infty} du \, e^{-u^2} \, [H_k(u)]^2 = 2^k \Gamma(k+1) \sqrt{\pi}$. This gives the bulk density,
\bea \label{rho_bullk_S}
{\rho_k^{\rm bulk} (r,\theta,N) \approx  \frac{1}{\pi } {\cal I}_{\sqrt{\lambda_-(k)})<z< \sqrt{\lambda_+(k)}} \;}\;,
\eea
where ${\cal I}_{\sqrt{\lambda_-(k)})<z< \sqrt{\lambda_+(k)}}$ is an indicator function that takes value $1$ if the inequality in the subscript is satisfied and $0$ otherwise.  \\

Let us summarise the results of above Sec~\ref{ssec:bulk}. The total density is obtained by summing over all the bands below the Fermi energy and is given by its large $N$ 
scaling form,
\bea \label{rho_scaling_S}
\rho(r,\theta,N) \sim f\left( \frac{r}{\sqrt{N}}\right) \;, 
\eea 
where the scaling function $f(z)$ is given by,
\bea \label{scaling_f_S}
f(z) = \frac{1}{\pi} \sum_{k=0}^{k^*} {\cal I}_{\sqrt{\lambda_-(k)}<z<\sqrt{\lambda_+(k)}} \;,
\eea
and $\lambda_{\pm}(k)$ is given in Eq. \ref{lpmk_S}. One can check that $f(z)$ in Eq. \ref{scaling_f_S} is normalized, i.e. $2 \pi \int_0^\infty f(z) \, z\, dz = 1$ upon using the definition of $\lambda_\pm(k)$ from Eq.~\ref{lpmk_S} and the relation in Eq. \ref{rel_mu_N2_S}.

Hence the limiting density has a compact single support over $\sqrt{\lambda_-(0)} < z < \sqrt{\lambda_+(0)} $. For $k^*=0$, it is just a simple flat density over this support. However, for $k^*>0$, the density has a nontrivial layered shape. For example, for $k^*=1$, the density is given by (see Fig.~\ref{figcomp_S}),
{\bea \label{density_kstar_1_S}
f(z) = 
\begin{cases}
& 0  \;\;, \; z < \sqrt{\lambda_-(0)}\\
& \\
& \dfrac{1}{\pi} \;\;, \; \sqrt{\lambda_-(0)} < z < \sqrt{\lambda_-(1)}\\
& \\
& \dfrac{2}{\pi} \;\;,\; \sqrt{\lambda_-(1)}< z< \sqrt{\lambda_+(1)} \\
& \\
& \dfrac{1}{\pi} \,\;, \; \sqrt{\lambda_+(1)} < z < \sqrt{\lambda_+(0)} \\ 
& \\
& 0  \;\;, \;  z >\sqrt{\lambda_+(0)} \;. \\ 
\end{cases}
\eea}

\subsection{Edges}
\label{ssec:edge}

In contrast if $z$ is close to one of the two edges, say the left edge $\sqrt{\lambda_-(k)}$, we can estimate the limiting form of the edge density when $N \to \infty$ from the same expression in Eq.~\ref{rho_k3_S}. For this, we set,
\bea \label{scal_var_S}
z^2 = \lambda_-(k) + \frac{\sqrt{2 \lambda_-(k)}}{\sqrt{N}} u \;, 
\eea
where $u \sim O(1)$. In this case, the lower limit in the integral in Eq.~\ref{rho_k3_S} becomes $a_-(k) \approx -u$ (with $u$ measuring the scaled distance from the left edge), while the upper limit still approaches to $+ \infty$ as $N \to \infty$. Hence we get,
\bea \label{rho_edge_S}
\rho_k^{\rm edge}(r,\theta,N) \to  f^{\rm edge}_k(u) 
\eea
where
\bea \label{rho_edge_S_e}
f_k^{\rm edge}(u) = \frac{2^{-k}}{\pi^{3/2} \Gamma(k+1)} \int_{-u}^{\infty} dx \, e^{-x^2} \, [H_k(x)]^2 \nonumber \\ \,\,
\eea
and we recall that,
\bea \label{exp_t_S}
u = \sqrt{\frac{N}{2 \lambda_-(k)}}\left(\frac{r^2}{N} - \lambda_-(k)\right) \;.
\eea
Note that when $u \to \infty$, $f_k^{\rm edge}(u) \to 1/\pi$, and the edge density matches smoothly with the bulk density.
\begin{figure*}[ht]
\includegraphics[width = 0.4\linewidth]{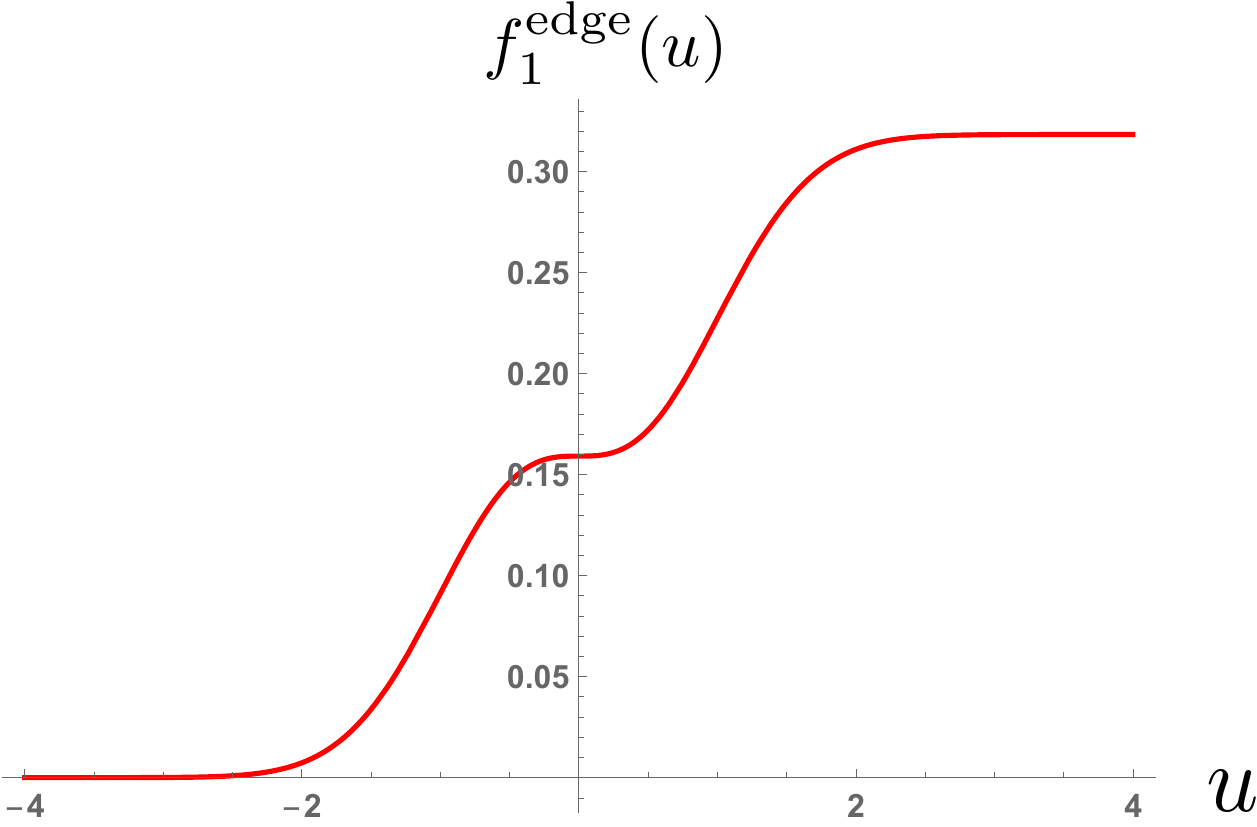}\,\,\quad
\includegraphics[width = 0.42\linewidth]{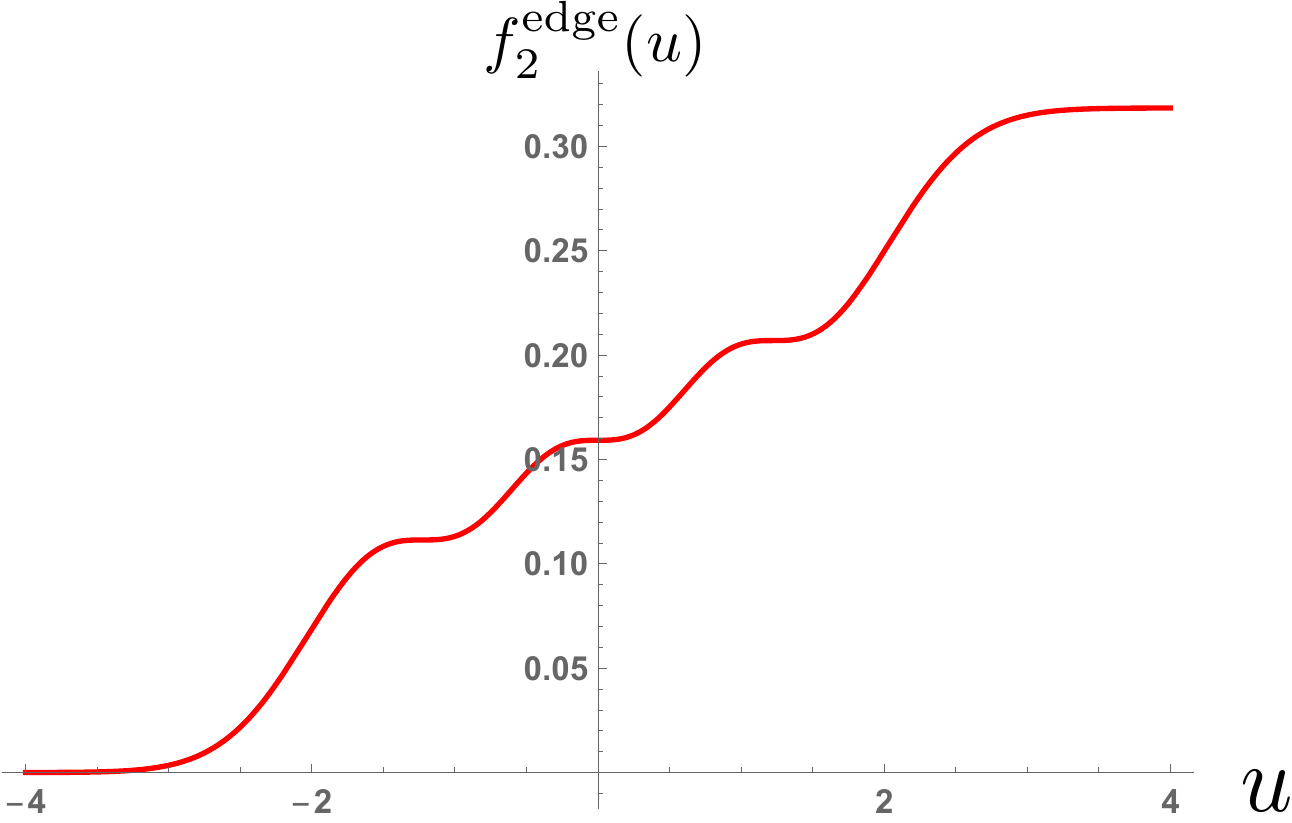}
\caption{The edge density scaling functions $f^{\rm edge}_k(u)$, i.e., Eq.~\ref{rho_edge_S} is plotted vs $u$ for $k=1$ (left panel) and $k=2$ (right panel). We see that the location of the kinks is at $H_k(u) = 0$.} \label{Fig_edge_S}
\end{figure*}
In Fig. \ref{Fig_edge_S} we have plotted the edge density functions $f^{\rm edge}_k(u)$ vs $u$ for $k=1$ and $k=2$. One sees from these figures that the scaling functions have kinks. For $k=1$, there is only one kink at $u=0$ while for $k=2$ there are two kinks. In general, for the $k^{\text{th}}$ band, the function $f_k(u)$ will have $k$ kinks as a function of $u$. The kinks occur when the derivative vanishes, i.e., $df^{\rm edge}_k(u)/du = 0$. By taking the derivative of Eq.~\ref{rho_edge_S}, we see that this happens when $H_k(-u) = 0$. Thus the locations of the kinks in the edge density of the $k^{\text{th}}$ band coincide with the zeroes of the $k^{\text{th}}$ Hermite polynomial. For instance, for $k=2$, the kinks are located at $u_1 = - 1/\sqrt{2}$ and $u_2 = +1/\sqrt{2}$. Note that the edge scaling function $f_k(u)$ is actually universal, i.e. independent of the system parameters $c$ and $M$ and depends only on the band label $k$.  The above non-trivial connection between $f^{\rm edge}_k (u)$ and Hermite polynomials naturally points to a possible connection to RMT, which we elucidate below. \\


\subsection{Edge density in the limit of high Landau levels ($k \gg 1$) and connection to Random Matrix Theory}

\begin{figure}[t]
\includegraphics[width = 1.0 \linewidth]{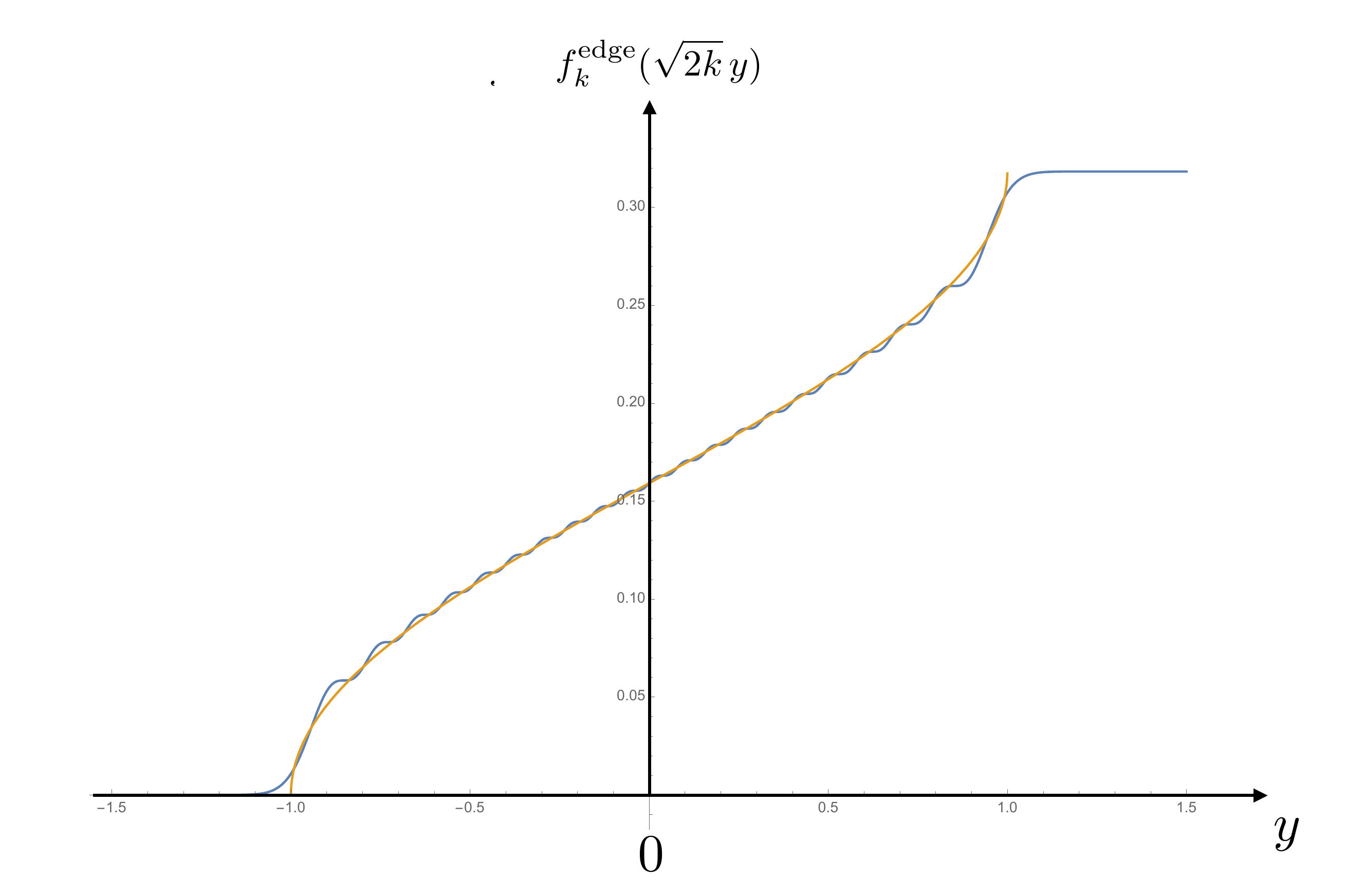}
\caption{Plot of $f_{k}^{\rm edge}(\sqrt{2k}\, y)$, with $f_{k}^{\rm
edge}(y)$ given in Eq.~\ref{rho_edge_S} as a function of $y$ for
$k=20$ (red solid line). The black-dashed line is the exact limiting
form given in Eq.~\ref{scaled_edge_4_S}. It should be noted that the
positions of kinks are at zeros of Hermite polynomial of degree
$k=20$. }\label{fig:S6}
\end{figure}

 It turns out that as $k \to \infty$, the edge profile $f_k^{\rm edge}(u)$ given in Eq.~\ref{rho_edge_S}, property shifted and scaled, has a nice limiting profile. This behaviour comes from the asymptotic behavior of the Hermite polynomials $H_k(u)$ in the limit of large $k$ (known as Plancherel-Rotach asymptotics). To obtain this limiting profile, we first set $u = \sqrt{2k} y$, with $y \sim O(1)$, and also perform the change of variable in Eq.~\ref{rho_edge_S}, $x = \sqrt{2k}(v-y)$. This leads to,
\begin{widetext}
\bea \label{scaled_edge_S}
 f_k^{\rm edge}(u = \sqrt{2k} y) = \frac{2^{-k}\sqrt{2 k}}{\pi^{3/2} \Gamma(k+1)} \int_{0}^{\infty} dv \, \left[e^{-k(v-y)^2} \, H_k(\sqrt{2k}(v-y))\right]^2  \;.
\eea
\end{widetext}
We can now use the Plancherel-Rotach asymptotic formula for Hermite polynomials,
\bea \label{Plancherel_1_S}
e^{- k X^2} H_k(\sqrt{2 k} X) &=& \left(\frac{2}{\pi}\right)^{1/4} \frac{2^{k/2}}{(1-X^2)^{1/4}} k^{-1/4} (k!)^{1/2}   \times  \nonumber \\ &&g_k(X) \left( 1 + {\cal O}\left(\frac{1}{k}\right) \right) \;, \; -1 < X  < 1 \nonumber \\
\eea
with
\bea 
g_k(X) = \cos\left( k X \sqrt{1-X^2} + (k + 1/2) \sin^{-1} X - k \pi/2\right)   \label{Plancherel_2_S}\nonumber \\
\eea

Inserting this expansion (Eq.~\ref{Plancherel_1_S} and Eq.~\ref{Plancherel_2_S}) with $X=v-y$ in Eq.~\ref{scaled_edge_S}, one finds,
\bea \label{scaled_edge_2_S}
 f_k^{\rm edge}(u = \sqrt{2k}\,  y) &\approx& \frac{2}{\pi^2} \int_{0}^\infty {\cal I}_{-1<v-y<1} \times  \nonumber \\ && \frac{1}{\sqrt{1-(v-y)^2}} \, g_k^2(v-y) \, dv\nonumber\\
\eea
where the indicator function comes from the fact that the asymptotic behavior in Eq.~\ref{Plancherel_1_S} and Eq.~\ref{Plancherel_2_S} holds only for $-1<X<1$, while it is sub-leading (in $k$) for $X$ outside the region. Due to the identity $\cos^2 x = 1/2 + \cos{(2 x)}/2$, one can replace $\left[g_{k}(v-y)\right]^2$, given in Eq.~\ref{Plancherel_2_S}, in the integral over $v$ in Eq.~\ref{scaled_edge_2_S} by $1/2$ (the remaining cosine being highly oscillating for large $k$ and thus subleading). Therefore we get,
\bea \label{scaled_edge_3_S}
f_k^{\rm edge}(u = \sqrt{2k}\,  y) \approx \frac{1}{\pi^2} \int_{\max(y-1,0)}^{\max(y+1,0)} \frac{dv}{\sqrt{1-(v-y)^2}} \nonumber \\
\eea
which finally yields (see also Fig. \ref{fig:S6}),
\begin{widetext}
\bea \label{scaled_edge_4_S}
\lim_{k \to \infty} f_k^{\rm edge}(u= \sqrt{2k}\,  y) =
\begin{cases}
& 0 \;, \;\hspace*{3.7cm} y < - 1 \\
& \\
& \dfrac{1}{\pi^2} \left( \dfrac{\pi}{2} + \sin^{-1}(y)\right) \;, \; -1 < y < 1 \\
& \\
& \frac{1}{\pi} \;, \; \;\hspace*{3.7cm}y > 1 \;.
\end{cases}
\eea
\end{widetext}
Close to $u = \pm \sqrt{2 k}$ there is an interesting edge region, of width $O(k^{-1/6})$ where the density is described by Airy functions, very similar to the well known ``Tracy-Widom'' regime at the edge of the Wigner semi-circle in RMT belonging to the Gaussian Unitary Ensemble (GUE). This is somehow expected given the square-root singularity near the edges $u = \pm \sqrt{2k}$ of the limiting profile given in Eq.~\ref{scaled_edge_4_S}. This edge behavior can be derived from Eq.~\ref{rho_edge_S} by using the asymptotic behavior of the Hermite polynomial $H_{k}(u)$ near $u = \sqrt{2 k}$ where the Hermite polynomial becomes an Airy function. One finds (for large $k$), setting $u=-\sqrt{2k} + \frac{w}{\sqrt{2} k^{1/6}}$ with $w = O(1)$
\bea \label{Airy_asympt_S}
f_k\left(-\sqrt{2k} + \frac{w}{\sqrt{2} k^{1/6}}\right) \sim \frac{1}{k^{1/3}} {\cal F}(w) 
\eea
where
\bea \label{Airy_asympt_S}
{\cal F}(w) = \frac{1}{\pi} \int_0^\infty {\rm Ai}^2(v-w) \, dv &=& \frac{1}{\pi} \big( [{\rm Ai}'(-w)]^2 \nonumber \\ &+& w {\rm Ai}^2(-w) \big)\nonumber \\
\eea
and ${\rm Ai}(z)$ denotes the standard Airy function. Note that a similar computation could be carried out for the kernel which would lead (on the real line at least) to the well known Airy kernel.

\section{Emergence of new droplet as one crosses critical lines in $(M,c)$ plane}
\label{sec:tw}

\begin{figure}[t]
\includegraphics[width = 1.0\linewidth]{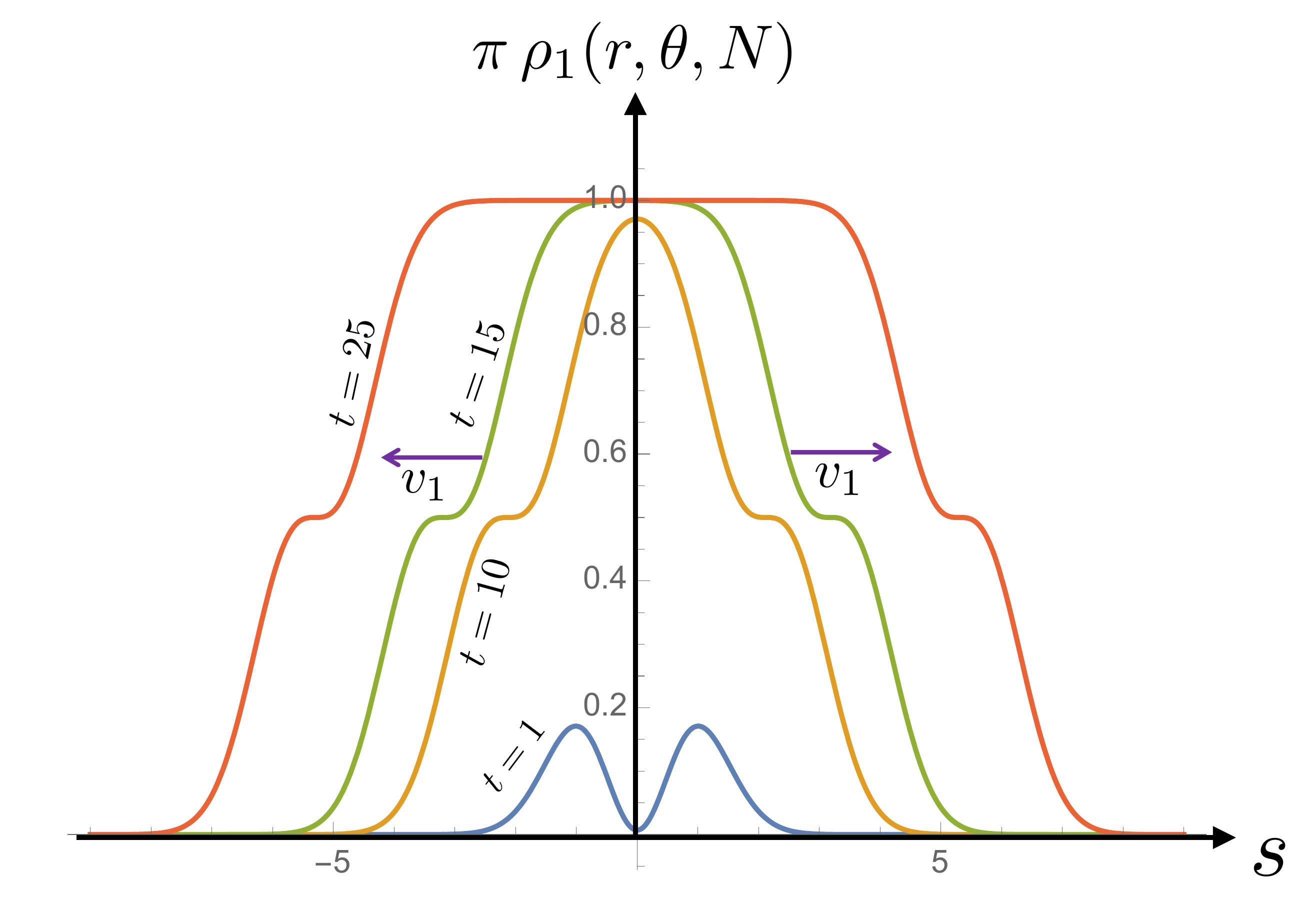}
\caption{Plot of the density profile $\, \pi \,\rho_1(r,\theta,N)$ as given in Eq.~\ref{delta_rho_6_S} for $M=5$, as a function of the scaled distance $s$ -- we recall that $r = z \sqrt{N} \approx \sqrt{N}(c_1/M)^{1/4} + s/\sqrt{2}$ -- for different increasing values of $t = 1, 10, 15$ and $25$ (from bottom to top). As $t$ increases, the scaled density $ \pi \,\rho_1(r,\theta,N)$ approaches the constant value $1$ for $|s| < v_1 t$ and decays rapidly to $0$ for $|s| > v_1 t$. The front separating the constant density $1/\pi$ and the zero-density outside ``moves with a constant speed $v_1$'' with increasing $t$, reminiscent of a remarkable travelling front structure.}\label{Fig_critical_S}
\end{figure}

We want to look at the phase diagram in the $(M,c)$ plane and ask, when $k^*$ changes from $k^*=0$ to $k^*=1$ (which means a new band is included below the Fermi energy), how does the density profile change from one layered structure to two layered structure. We have already seen that just when one crosses this critical line $c=c_1(M)\equiv c_1$, the second layer appears on top of the first layer. In this subsection, we describe the density profile of this emerging blob in the second layer for $c$ slightly below $c_1$ for fixed $4<M<12$ (see Fig.~\ref{fig:cmp_S}), where $c_1$ is given in Eq. \ref{c1M_S}. We therefore set 
\begin{equation} \label{def_Delta_S}
c = c_1 - \Delta \;\; {\rm where} \;\; 0 < \Delta \ll 1 \;. 
\end{equation}
For each point in the $(M,c)$ plane, $\mu$ is uniquely determined from Eq. \ref{rel_mu_N2_S}. Therefore as we change the value of $c$ from $c_1$ to $c_1 - \Delta$, the value of $\mu$ also changes from $\mu = 2 + \sqrt{c_1 M}$ to 
\bea \label{def_delta0_S}
\mu = 2 + \sqrt{c_1 M} + \delta \;,
\eea   
where $\delta \ll 1$. Inserting this value of $\mu$ in Eq. \ref{rel_mu_N2_S} and expanding for small $\delta$ gives a relation between $\Delta$ and $\delta$
\bea \label{rel_delta_Delta_S}
\Delta = \frac{(M^2-16)^{3/2}}{64\,\sqrt{2}} \sqrt{\delta} \;.
\eea 
Therefore, one just has one single control parameter $\delta$ describing the location of the system in the phase diagram, in the vicinity
of the critical line $c=c_1(M)$. We now want to see how the density changes as we vary $\delta$.

We start with the formula for the density in Eq.~\ref{gen_density_S}. When $k^*$ increases from $0$ to $1$, the additional density in the second layer is given by,
\bea \label{delta_rho_S}
\rho_1(r,\theta,N) &=& \frac{e^{-r^2}}{\pi}  \sum_{l = l_-(1)}^{l_+(1)} \frac{r^{2 \sqrt{\gamma + l^2}}}{\Gamma(\sqrt{\gamma + l^2}+2)} \left[L_1^{\sqrt{\gamma+l^2}}(r^2)\right]^2  \nonumber \\ 
\eea 
with
\bea \label{delta_rho_S_e}
L_1^{\sqrt{\gamma+l^2}}(r^2) = 1 + \sqrt{\gamma+l^2}-r^2 
\eea 

where, 
\bea \label{lpm1_S}
l_{\pm}(1) = \frac{\mu-2 \pm \sqrt{(\mu - 2)^2 - c_1 M}}{M} \, N \;.
\eea
We can rewrite Eq.~\ref{lpm1_S} using Eq.~\ref{def_delta0_S} (to leading order in $\delta$ for small $\delta$) as, 
\bea \label{lpm1_2_S}
l_{\pm}(1) \approx \lambda_{\pm} \,N, \quad \lambda_{\pm} = \left(\sqrt{\frac{c_1}{M}} \pm v_1 \sqrt{\delta} \right) 
\eea
with 
\bea \label{lpm1_2_S}
v_1 = \frac{\sqrt{2}}{M} (c_1 M)^{1/4} \;.
\eea
In real space, the second layer of the macroscopic density appears over the scaled region, 
\bea \label{interval_S}
\sqrt{\frac{c_1}{M}} - v_1 \sqrt{\delta} < z^2 < \sqrt{\frac{c_1}{M}} + v_1 \sqrt{\delta} \;,
\eea 
where $z=r/\sqrt{N}$. Therefore, the center of the second layer is located at $z_c = (c_1/M)^{1/4}$ and we want to provide a scaling description of this density in the second layer just after its appearance, i.e., in the limit $\delta \to 0$. Hence we set, 
\begin{eqnarray} \label{def_epsilon_S}
z^2 = \sqrt{\frac{c_1}{M}} + \epsilon \;, 
\eea
where $\epsilon$ is proportional to the distance from the center of the second layer. Thus the density is just a function of $\epsilon$ and $\delta$ in the vicinity of the critical line $c=c_1(M)$ and below we work out the dependence of the density on these two parameters in the large $N$ limit.

To analyse the density Eq.~\ref{delta_rho_S} in the limit of large $N$, we set $\gamma = c_1 \, N$ and we introduce $l = x \, N$ so that the sum over $l$ can be replaced by an integral over $x$ leading to, 
\begin{widetext}
\bea \label{delta_rho_2_S}
\rho_1(r,\theta,N) \approx \frac{e^{-r^2}}{\pi} \int_{\lambda_-}^{\lambda_+} dx \frac{r^{2\sqrt{c N + N^2 x^2}}}{\Gamma(\sqrt{cN+N^2x^2}+2)}\left[  1 + \sqrt{c N+N^2 x ^2}-r^2\right]^2 \approx  \frac{e^{-r^2}}{\pi} \int_{\lambda_-}^{\lambda_+} dx \frac{r^{2N x}}{\Gamma(Nx+2)}\left(N\,x - r^2\right)^2 \nonumber \\
\eea
\end{widetext}
 where we kept the leading term in the arguments for large $N$. We can now approximate the Gamma function by the Stirling's formula, leading to,
\begin{widetext}
\bea  \label{delta_rho_3_S}
\rho_1(r = z \sqrt{N},\theta,N) \approx \frac{e^{-N z^2}}{\pi \sqrt{2 \pi N}} \int_{\lambda_-}^{\lambda_+} \frac{dx}{x^{3/2}}\, e^{2 N x \ln (z \sqrt{N}) - N\,x\ln x + N\,x}(Nx-N z^2)^2 \;.
\eea 
\end{widetext}
We now substitute $z^2 =  \sqrt{\frac{c_1}{M}} + \epsilon$ from Eq.~\ref{def_epsilon_S} and make the change of variable $x = \sqrt{c_1/M} + v$. Since $|v| < v_1 \sqrt{\delta}$, we can expand the integrand for small $v$ and retain only up to $O(v^2)$ terms inside the exponential. After straightforward algebra, one obtains,
\begin{widetext}
\bea \label{delta_rho_4_S}
\rho_1(r = z \sqrt{N},\theta, N) \approx \frac{1}{\pi \sqrt{2 \pi N}} \left( \frac{M}{c_1}\right)^{3/4} \int_{-v_1 \sqrt{\delta}}^{+v_1 \sqrt{\delta}} dv \, e^{-{N}\sqrt{{\frac{M}{4c_1}}}(v-\epsilon)^2}N^2(v-\epsilon)^2 \;.
\eea
\end{widetext}
In order that this integral is of order $O(1)$, we see that we need to scale $\sqrt{\delta} \sim t/\sqrt{N}$,  $\epsilon \sim s/\sqrt{N}$ where $t>0$ as well as $s$ are both of order $O(1)$. Making the change of variable $\tilde{w} = [N\,/2]^{1/2}(M/c_1)^{1/4}(v-\epsilon)$ in Eq.~\ref{delta_rho_4_S}, we get,
\bea \label{delta_rho_5_S}
\rho_1(r= z \sqrt{N},\theta, N) \approx \frac{2}{N} \frac{1}{\pi^{3/2}} \int_{\tilde{w}_-}^{\tilde{w}_+} d\tilde{w} \, \tilde{w}^2 \, e^{-\tilde{w}^2} 
\eea
with 
\bea \label{delta_rho_5_S_e}
\tilde{w}_{\pm} = \sqrt{\frac{N}{2}} \left( \frac{M}{c_1}\right)^{1/4}(\pm v_1 \sqrt{\delta} - \epsilon) \;.
\eea
In order that the integral remains of order $O(1)$ in the large $N$ limit, we see that both $\sqrt{\delta}$ and $\epsilon$ should scale as $O(1/\sqrt{N})$. We therefore set,
\bea \label{def_delta_S}
\sqrt{\delta} = \sqrt{\frac{2}{{N}}} \left({\frac{c_1}{M}}\right)^{1/4}\,t \;\;\;\; {\rm and} \;\;\;\; \epsilon =  \sqrt{\frac{2}{{N}}} \left({\frac{c_1}{M}}\right)^{1/4}\,s \;,\nonumber \\
\eea
where $t$ and $s$ are both of order $O(1)$. Therefore, the density in the large $N$ limit, a function of the original variables $\epsilon$ and $\delta$,  
can be re-paramaterized in terms of the scaled variables $s$ and $t$ given in Eq. \ref{def_delta_S} 
\bea \label{delta_rho_6_S}
\rho_1(r = z \sqrt{N},\theta, N) \approx \frac{1}{\pi } \left[ F_1(s+v_1 t )    -  F_1(s-v_1 t)   \right] \;, \nonumber \\
\eea
where $v_1 = \frac{\sqrt{2}}{M}(c_1 M)^{1/4}$ and 
\bea \label{F_scaling_S}
F_1(z) = \frac{2}{\sqrt{\pi}}\,\int_0^z d \tilde w \, \tilde w^2 e^{-\tilde w^2} = \frac{1}{2} \left[{\rm erf}(z) - \frac{2}{\sqrt{\pi}} z \, e^{-z^2} \right] \;.\nonumber \\
\eea

Note that the scaled variables $t$ and $s$ can be expressed in terms of $\Delta = c_1 - c$ (which measures the location the distance in the phase diagram with respect to the critical line $c=c_1(M)$) and the variable $z = r/\sqrt{N}$ where $r$ measures the distance from the center of the trap. The first relation can be obtained by eliminating $\delta$ between Eqs. \ref{rel_delta_Delta_S} and \ref{def_delta_S} 
\bea \label{rel_t_Deltac_S}
t = \left({\frac{M}{4c_1}}\right)^{1/4}\frac{64\, \sqrt{2}}{(M^2-16)^{3/2}} \left[c_1-c\right]\sqrt{N} \;.
\eea 
Similarly the second relation is obtained by substituting $\epsilon = z^2 - \sqrt{c_1/M}$ in Eq. \ref{def_delta_S}. This gives
\bea \label{rel_s_z_S}
s =  \left({\frac{M}{4c_1}}\right)^{1/4}\left(z^2 - \frac{c_1}{M} \right) \sqrt{N} \;,
\eea
where $z = r/\sqrt{N}$. 

Interestingly, the scaled density profile in Eq. \ref{delta_rho_6_S} has an interesting traveling front structure. To see this, we consider the density as a function of $s$, for a fixed $t$. The density decays to $0$ very rapidly as $|s| \gg v_1 t$ (see Fig. \ref{Fig_critical_S}). Therefore the two edges of this profile move ``ballistically'' with increasing $t$ with a ``speed'' given by $v_1$. If we interpret $t$ as a ``time'', then at late times, the density profile develops a traveling front structure with velocity $v_1$ and the width across the front remains of $O(1)$ as $t$ increases.   
For large $t$, the density has a constant value $\simeq 1/\pi$ for all $|s| < v_1 t$ (see Fig.~\ref{Fig_critical_S}). Finally, the speed $v_1$ is given by $v_1 = \sqrt{2}/M (c_1 M)^{1/4}$ can be expressed in terms of $M$, by using the expression for $c_1(M)$ in Eq. \ref{c1M_S}.

Note that here we analysed the density profile near the transition from $k^*=0$ to $k^*=1$ where the second layer just appears over the first layer. One can do a similar analysis for the transition form $k^*=n-1$ to $k^*=n$ across the critical line $c=c_n(M)$ for any $n \geq 1$. We do not repeat the analysis here but it is easy to show that the scaled density will again be given by the difference of two functions, as in the $k=1$ case in Eq.~\ref{delta_rho_6_S}, 
\bea \label{final_rhok_S}
\rho_n(r = z \sqrt{N}) \approx \frac{1}{\pi} \left[ F_n(s+v_n t )    -  F_n(s-v_n t)   \right] \;,
\eea
where the speed $v_n$ can be computed from the critical curve $c=c_n(M)$ and the scaling function $F_n(z)$ is given, up to an overall constant by, 
\bea \label{Fn_S}
F_n(z) \propto \int_0^z d \tilde w\, \left[H_n(\tilde w)\right]^2 \, e^{-\tilde w^2} \;,
\eea
where $H_n(\tilde w)$ is the Hermite polynomial of degree $n$.


\begin{thebibliography}{52}
\bibitem{footnote_scaled_s} We recall that the relation between $r$ and $s$ reads
$r = z \sqrt{N} \approx ({c_1/M})^{1/4} \sqrt{N} + {s}/{\sqrt{2}}$ for large $N$.
\end{thebibliography}
\end{document}